\documentclass[letterpaper,twocolumn,10pt]{article}
\usepackage{usenix}

\usepackage{tikz}
\usepackage{amsmath}

\usepackage[]{todonotes}
\usepackage{xcolor}
\usepackage{url}
\usepackage[most]{tcolorbox}
\usepackage[nolist]{acronym}
\usepackage{booktabs}

\usepackage{xurl}

\usepackage[utf8]{inputenc}
\usepackage{amsmath, amssymb}
\usepackage{fontawesome5}
\usetikzlibrary{shapes.geometric, arrows.meta, positioning, fit, backgrounds, calc, shadows, patterns, shadows.blur}

\newcommand{\mypar}[1]{\smallskip\noindent\textbf{#1.}}

\newcommand{\myemphlist}[2]{\smallskip\noindent(#1)~\emph{#2.}}

\definecolor{mylightblue}{HTML}{D9EAFD}

\newtcolorbox[auto counter, number within=section]{marcosboxgreyopen}{%
  colback=gray!10, % Background color
  colframe=black, % Border color
  toprule=-0.25mm, % Remove top border
  bottomrule=-0.25mm, % Remove bottom border
  leftrule=1mm, % Add left border
  rightrule=-0.25mm, % Remove right border
  left=0.1mm, right=0.1mm, top=0.1mm, bottom=0.1mm, % Padding
  enhanced,
  arc=0mm,
  breakable = true,
}

\newtcolorbox[auto counter, number within=section]{marcosboxgreyclosed}{%
  colback=gray!10, % Background color
  colframe=black, % Border color
  toprule=-0.25mm, % Remove top border
  bottomrule=-0.25mm, % Remove bottom border
  leftrule=-0.25mm, % Add left border
  rightrule=1mm, % Remove right border
  left=0.1mm, right=0.1mm, top=0.1mm, bottom=0.1mm, % Padding
  enhanced,
  arc=0mm,
  breakable = true,
}

\newtcolorbox[auto counter, number within=section]{marcosboxredopen}{%
  colback=gray!80!red!10, % Background color
  colframe=red!40!black, % Border color
  toprule=-0.25mm, % Remove top border
  bottomrule=-0.25mm, % Remove bottom border
  leftrule=1mm, % Add left border
  rightrule=-0.25mm, % Remove right border
  left=0.1mm, right=0.1mm, top=0.1mm, bottom=0.1mm, % Padding
  enhanced,
  arc=0mm,
  breakable = true,
}

\newtcolorbox[auto counter, number within=section]{marcosboxblueopen}{%
  colback=gray!60!blue!10, % Background color
  colframe=blue!30!black, % Border color
  toprule=-0.25mm, % Remove top border
  bottomrule=-0.25mm, % Remove bottom border
  leftrule=1mm, % Add left border
  rightrule=-0.25mm, % Remove right border
  left=0.1mm, right=0.1mm, top=0.1mm, bottom=0.1mm, % Padding
  enhanced,
  arc=0mm,
  breakable = true,
}

\usepackage{minted}
\setminted{
  bgcolor=gray!8,
  frame=lines,
  fontsize=\small,
}

\usepackage[most]{tcolorbox}
\usepackage{enumitem}

\newcommand{\circnum}[1]{%
\tikz[baseline={([yshift=-3.2pt]C.center)}]\node[%  <-- Adjusted baseline to center
  circle, 
  fill=black, 
  inner sep=0pt,        % Changed to 0pt for more control
  minimum size=11pt,    % Consistent diameter
  font=\color{white}\scriptsize\bfseries % Moved color/font here
] (C) {#1};%
}

\newcounter{circsubsub}[subsection]

\newcommand{\circsubsubsection}[1]{%
  \refstepcounter{circsubsub}%
  \subsubsection*{\circnum{\thecircsubsub}\quad #1}
}

\usepackage{xspace}
\newcommand{\ri}{\textsc{Re-Moat}\xspace}
\newcommand{\sys}{\textsc{Moat}\xspace}

\newcommand{\pair}[2]{#1\,|\,#2}

\usepackage{fontawesome5}

\begin{acronym}
  \acro{ML}{Machine Learning}
  \acro{AI}{Artificial Intelligence}
  \acro{DL}{Deep Learning}

  \acro{NN}{Neural Network}
  \acrodefplural{NN}{Neural Networks}

  \acro{LLM}{Large Language Model}
  \acrodefplural{LLM}{Large Language Models}

  \acro{LSM}{Linux Security Module}
  
  \acro{PoC}{Proof-of-Concept}
  \acrodefplural{PoC}{Proofs-of-Concept}
  \acro{HIDS}{Host-based Intrusion Detection Systems}
\end{acronym}

\begin{document}
%-------------------------------------------------------------------------------

%don't want date printed
\date{}

\title{\Large \bf 
Lifecycle-Aware Dynamic Analysis for Secure ML Model Execution \\
}

%for single author (just remove % characters)
\renewcommand{\thefootnote}{\fnsymbol{footnote}}

\author{
{\rm Gabriele Digregorio\footnotemark[1]}\\
Politecnico di Milano
\and
{\rm Marco Di Gennaro\footnotemark[1]}\\
Politecnico di Milano
\and
{\rm Francesco Pastore}\\
Politecnico di Milano
\and
{\rm Stefano Zanero}\\
Politecnico di Milano
\and
{\rm Stefano Longari}\\
Politecnico di Milano
\and
{\rm Michele Carminati}\\
Politecnico di Milano
}

\maketitle

\footnotetext[1]{These authors equally contributed to this work.}

\renewcommand{\thefootnote}{\arabic{footnote}}
\setcounter{footnote}{0}

%-------------------------------------------------------------------------------
\begin{abstract}
The growing reliance on pre-trained \ac{ML} models has introduced new attack surfaces. Recent vulnerabilities demonstrate that malicious behavior can be embedded within model artifacts, often bypassing existing defenses. 
Current model-scanning solutions primarily rely on static, format-specific rules or known attack signatures, which limit their ability to generalize across frameworks and to detect novel exploitation paths.
In contrast, we propose a solution that focuses on the effects an attack has on the host system executing the model and builds on foundational intuitions about \ac{ML} model execution.
In particular, we observe that \ac{ML} models operate within well-defined lifecycle phases and that, within each phase, interactions with the host system are highly structured and predictable.
We translate these intuitions into \sys, a dynamic lifecycle-aware approach for securing \ac{ML} model execution, and instantiate this design in \ri, our reference implementation.
We evaluate \ri across multiple \ac{ML} frameworks using 77{,}974 real-world model artifacts from the Hugging Face Hub, 31 \acp{PoC} from CVEs, and 334 models from a state-of-the-art dataset, and compare it against state-of-the-art model-scanning solutions. Our results show that our approach detects all evaluated attack classes while maintaining a close-to-zero false-positive rate, validating our intuitions and motivating dynamic analysis for securing \ac{ML} model execution.
\end{abstract}

\section{Introduction}\label{sec:intro}
The adoption of \acf{ML} is growing rapidly, driven by platforms for sharing pre-trained models~\cite{HuggingFaceHub, KaggleModels, TensorFlowHub, PyTorchHub}. This development mirrors traditional software engineering, where open-source repositories facilitate the reuse and adaptation of code. Similarly, today’s \ac{ML} ecosystem relies on the sharing and reuse of model artifacts.

Despite the clear advantages, this transition has raised concerns about the security implications of loading and executing \ac{ML} models obtained from public repositories or shared through other channels. Recent research has highlighted the risks associated with running untrusted models~\cite{jiang2022empirical,jiang2023empirical,meiklejohn2025machine,zhu2025my,digregorio2026onthe}, while major model-sharing platforms (e.g., the Hugging Face Hub) have increasingly expanded their set of built-in or third-party \ac{ML} model artifact scanners to mitigate these threats~\cite{ClamAV,VirusTotal,HuggingFacePickleSecurity,JFrog,ModelScanProtectAI}. The risks posed by malicious models range from unauthorized file access and data exfiltration to arbitrary code execution~\cite{CVE-2024-3660,CVE-2025-1550,CVE-2025-8747,CVE-2025-9905,CVE-2025-9906,CVE-2025-12058,CVE-2025-32434,CVE-2025-49655}. 
As a result, downloading and running an untrusted \ac{ML} model increasingly resembles executing untrusted software at large.

Although malicious software detection is a mature and well-studied area, the problem of \ac{ML} model scanning remains underexplored. Existing model scanners provide limited and fragile protection. Most tools~\cite{pickleball,flickling} either focus narrowly on Pickle-based artifacts or rely on static, signature-driven rules that struggle to generalize across different file formats (e.g., \texttt{.pt}, \texttt{.keras}), \ac{ML} frameworks (e.g., Pytorch, TensorFlow, and Keras), and evolving attack techniques. Consequently, these scanners are largely reactive, detecting known attack patterns while failing to identify new attacks that exploit previously unexplored techniques or abuse framework features~\cite{zhu2025my}. Empirical studies~\cite{digregorio2026onthe,casey2024large} further show that current solutions suffer from false negatives and false positives in real-world deployments.

As a result, recent works~\cite{digregorio2026onthe} have emphasized the need to draw inspiration from decades of research in malware detection when addressing \ac{ML} model security. At the same time, it is crucial to account for the distinctive properties that distinguish \ac{ML} models from generic software. 

\mypar{Our Intuition} Malicious \ac{ML} models can be effectively identified by the effects they induce on the host system at runtime (dynamic analysis), leveraging the characteristic properties of \ac{ML} models. This runtime perspective can complement static analysis, addressing cases that signature- and format-specific tools miss.
In particular, \ac{ML} model execution is organized around a small number of well-defined lifecycle phases (e.g., loading, inference, and training), and different \ac{ML} artifacts should not be treated as entirely distinct software instances. Rather, they represent variations within the capability space exposed by a specific framework. As a consequence, the set of interactions that model execution is \emph{expected} to have with the host system is narrow and highly predictable, making deviations a strong signal of potentially malicious behavior (allowlisting approach).

\mypar{Our Approach} We translate these intuitions into \sys, a lifecycle-aware approach for securing \ac{ML} model execution by enforcing execution boundaries over system-level interactions. \sys is structured around four components: (i) an \emph{action abstraction} that maps low-level execution events to security-relevant interactions with the host system; (ii) the \emph{execution boundaries definition}, which specifies the actions expected during each lifecycle phase; (iii) an \emph{orchestrator}, which coordinates the analysis across different lifecycle phases and (iv) a \emph{tracer} that observes system-level interactions at runtime and detects boundary violations. 
We realize this approach in \ri, a reference implementation that instantiates these components by tracing system calls and mapping them to security-relevant actions.

\mypar{Evaluation} 
We evaluate \ri across multiple frameworks and lifecycle phases against 23 \acp{PoC} from 8 public CVEs, 4 \acp{PoC} derived from a vulnerability not yet publicly disclosed at the time of analysis, 77{,}974 real-world models downloaded from the Hugging Face Hub, and 4 \acp{PoC} from a state-of-the-art attack~\cite{zhu2025my}. We further compare \ri with state-of-the-art approaches on 334 models from a dataset used in prior work~\cite{pickleball}. Under our threat model, which targets model behaviors that aim to compromise the host system, \ri exhibits no observed false negatives or false positives (0\%) on the evaluated datasets, thereby improving detection coverage while reducing spurious alerts compared to existing approaches.
Notably, under the same threat model, \ri outperforms approaches specifically tailored to Pickle artifacts when evaluated on Pickle-based models, while remaining agnostic to the underlying model format.
These results demonstrate a robust and generalizable defense for \ac{ML} model execution across diverse attack patterns, ranging from unsafe serialization to code reuse and framework-specific vulnerabilities, and across different model formats, from framework-specific representations to standard serialization formats such as Pickle.

\mypar{Contributions} We make the following contributions:

\begin{itemize}[leftmargin=*]
    \item We derive foundational intuitions under which the structured and predictable execution of \ac{ML} models enables the definition of execution boundaries. These boundaries define which interactions with the host system are \emph{expected}, allowing deviations to be systematically identified as suspicious.

    \item We translate these intuitions into \sys, a lifecycle-aware approach for securing model execution by enforcing execution boundaries over interactions with the host system.

    \item We design and implement \ri, a syscall-based reference implementation of \sys that traces system calls at runtime and maps them to security-relevant actions. \ri supports multiple \ac{ML} frameworks and file formats.

    \item We empirically validate our intuitions by evaluating \ri on 77{,}974 real-world models, 31 \acp{PoC}, and 334 models from a state-of-the-art dataset.
    
\end{itemize}

\mypar{Open Science} After peer-reviewed publication, we will provide all artifacts necessary to reproduce our findings.

\section{Background}\label{sec:background}
We provide background on \ac{ML} model sharing (or \ac{ML} model \textit{supply chain}), the approaches adopted by different \ac{ML} frameworks, and the associated risks. 

\subsection{ML Models' Lifecycle}\label{subsec:model-lifecycle}
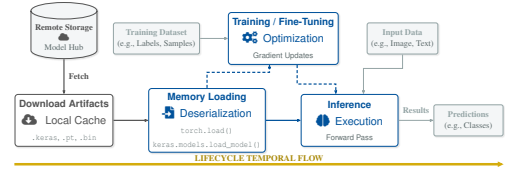
\begin{figure}[t]
    \centering
    \resizebox{0.8\linewidth}{!}{% --- Professional Color Palette ---
\definecolor{colCore}{HTML}{00529B}      % Professional Blue
\definecolor{infraGrey}{HTML}{555555}    % Dark Grey
\definecolor{dataGrey}{HTML}{95A5A6}     % Silver/Light Grey
\definecolor{dataFill}{HTML}{F4F6F7}     % Very light fill
\definecolor{timelineYellow}{HTML}{D4AC0D} 
\definecolor{textGray}{HTML}{5F6C6D}

\newcommand{\subText}[1]{{\scriptsize\color{textGray} #1}}

\begin{tikzpicture}[
    font=\sffamily,
    transform shape,
    scale=1.2,
    node distance=0.8cm and 1.0cm,  % Reduced from 1.2cm and 1.8cm
    % --- Styles ---
    processCore/.style={
        rectangle, rounded corners=3pt,
        draw=colCore, line width=1pt, fill=white,
        minimum width=2.8cm, minimum height=1.5cm,  % Reduced from 3.5cm x 1.8cm
        align=center, text=colCore,
        blur shadow={shadow blur steps=5, shadow xshift=1pt, shadow yshift=-1pt, shadow opacity=20}
    },
    processInfra/.style={
        rectangle, rounded corners=3pt,
        draw=infraGrey, line width=1pt, fill=white,
        minimum width=2.8cm, minimum height=1.5cm,  % Reduced
        align=center, text=infraGrey,
        blur shadow={shadow blur steps=5, shadow xshift=1pt, shadow yshift=-1pt, shadow opacity=20}
    },
    sourceInfra/.style={
        cylinder, shape border rotate=90, aspect=0.25,
        draw=infraGrey, line width=1pt, fill=dataFill,
        minimum width=1.8cm, minimum height=1.3cm,  % Reduced from 2.2cm x 1.6cm
        align=center, font=\scriptsize, text=infraGrey,
        blur shadow={shadow blur steps=5, shadow xshift=1pt, shadow yshift=-1pt, shadow opacity=20}
    },
    dataBlock/.style={
        rectangle, rounded corners=2pt,
        draw=dataGrey, line width=1pt, fill=dataFill,
        minimum width=2.0cm, minimum height=0.9cm,  % Reduced from 2.5cm x 1.1cm
        align=center, font=\scriptsize, text=dataGrey,
        blur shadow={shadow blur steps=5, shadow xshift=1pt, shadow yshift=-1pt, shadow opacity=15}
    },
    % --- Arrow Styles (consistent arrowhead size) ---
    myarrow/.style={
        >={Stealth[round, length=6pt, width=4pt]}  % Slightly smaller arrows
    },
    infraFlow/.style={
        myarrow, ->, line width=1.2pt,
        color=infraGrey
    },
    coreFlow/.style={
        myarrow, ->, line width=1.2pt,
        color=colCore
    },
    dataFlow/.style={
        myarrow, ->, line width=1pt,
        color=dataGrey
    },
    optFlow/.style={
        dashed, dash pattern=on 3pt off 2pt
    },
    pathLabel/.style={
        font=\scriptsize\bfseries\itshape, fill=white, inner sep=1pt, align=center, text=colCore
    }
]

% ==========================================================
% 1. NODES
% ==========================================================

% --- Main Pipeline Row (Bottom) ---
\node[processInfra] (Download) {
    \textbf{\small Download Artifacts}\\[2pt]
    \faCloudDownload* \enspace Local Cache\\[1pt]
    {\subText{\texttt{.keras}, \texttt{.pt}, \texttt{.bin}}}
};

% Remote Storage is ABOVE Download
\node[sourceInfra, above=1.0cm of Download] (Cloud) {
    \textbf{Remote Storage}\\[1pt]
    \faCloud\\
    {\subText{Model Hub}}
};

\node[processCore, right=of Download] (Load) {
    \textbf{\small Memory Loading}\\[2pt]
    \faFileImport \enspace Deserialization\\[1pt]
    {\subText{\texttt{torch.load()}}}\\
    {\subText{\texttt{keras.models.load\_model()}}}
};

\node[processCore, right=of Load] (Infer) {
    \textbf{\small Inference}\\[2pt]
    \faBrain \enspace Execution\\[1pt]
    {\subText{Forward Pass}}
};

\node[dataBlock, right=of Infer] (OutputData) {
    \textbf{Predictions}\\[1pt]
    {\subText{(e.g., Classes)}}
};

% --- Training Row (Top) ---
\node[processCore, above=1.6cm of $(Load.east)!0.5!(Infer.west)$] (Train) {
    \textbf{\small Training / Fine-Tuning}\\[2pt]
    \faCogs \enspace Optimization\\[1pt]
    {\subText{Gradient Updates}}
};

% Training Data (to the left of Train)
\node[dataBlock, left=0.8cm of Train] (TrainData) {
    \textbf{Training Dataset}\\[1pt]
    {\subText{(e.g., Labels, Samples)}}
};

% Input Data (to the right of Train)
\node[dataBlock, right=0.8cm of Train] (InputData) {
    \textbf{Input Data}\\[1pt]
    {\subText{(e.g., Image, Text)}}
};

% ==========================================================
% 2. CONNECTIONS
% ==========================================================

% --- Infrastructure Flow (Cloud above Download) ---
\draw[infraFlow] (Cloud) -- node[right, font=\scriptsize\bfseries, color=infraGrey] {Fetch} (Download);
\draw[infraFlow] (Download) -- (Load);

% --- Core Flow: Load to Infer (Solid) ---
\draw[coreFlow] (Load) -- node[midway, above=2pt, pathLabel] {} (Infer);

% --- Data Flows ---
\draw[dataFlow] (Infer) -- node[midway, above=2pt, font=\scriptsize\bfseries, color=dataGrey] {Results} (OutputData);

% Training Data -> Training Box
\draw[dataFlow] (TrainData.east) -- (Train.west);

% Input Data -> Inference Box (straight down then turn)
\draw[dataFlow] (InputData.south) -- ++(0, -0.4) -| ($(Infer.north) + (0.4, 0)$);

% --- Optional Detour: Load -> Train (Dashed, straight lines) ---
\draw[coreFlow, optFlow] 
    (Load.north) -- ++(0, 0.45) -| ($(Train.south) + (-0.4, 0)$);

% --- Fine-tuned Model: Train -> Infer (Dashed, straight lines) ---
\draw[coreFlow, optFlow] 
    ($(Train.south) + (0.4, 0)$) -- ++(0, -0.55) -| ($(Infer.north) + (-0.4, 0)$);

% ==========================================================
% 3. TIMELINE INDICATOR
% ==========================================================
\begin{scope}[on background layer]
    \coordinate (YRef) at ($(Infer.south) + (0, -0.5cm)$);
    \coordinate (TStart) at (Download.west |- YRef);
    \coordinate (TEnd)   at (OutputData.east |- YRef);

    \draw[line width=2.5pt, color=timelineYellow, -{Triangle[length=6pt, width=5pt]}] (TStart) -- (TEnd);
    \node[above=1pt, font=\bfseries\scriptsize\color{timelineYellow!80!black}, fill=white, inner sep=1pt] at ($(TStart)!0.5!(TEnd)$) {LIFECYCLE TEMPORAL FLOW};
\end{scope}

\end{tikzpicture}}
    \caption{Lifecycle of an \ac{ML} model, from artifact retrieval to prediction.}
    \label{fig:lifecycle}
\end{figure}

\ac{ML} models are typically used through a sequence of well-defined stages that constitute their lifecycle, as illustrated in Figure~\ref{fig:lifecycle}. In this work, we refer to these stages as \emph{lifecycle phases}, each representing a logically distinct execution stage in which the model and the \ac{ML} framework perform a specific, semantically coherent set of operations.
Across \ac{ML} frameworks, three primary lifecycle phases are commonly observed: \emph{loading}, \emph{inference}, and \emph{training}. The \emph{loading} phase involves reading a model artifact from persistent storage and reconstructing an in-memory model object for subsequent use. During \emph{inference}, the model processes inputs to produce predictions without modifying its parameters. The \emph{training} phase instead consists of updating the model’s parameters based on training data. 

\subsection{ML Model Sharing}\label{subsec:ml-model-sharing}
Instead of training models from scratch, practitioners increasingly download and reuse third-party \ac{ML} model artifacts from dedicated repositories (\textit{``model hubs''}) and framework-specific distribution channels~\cite{HuggingFaceHub,KaggleModels,TensorFlowHub,PyTorchHub}. This ecosystem resembles traditional software distribution. However, the primary shared object is often a \emph{model artifact} (e.g., \texttt{.pt}, \texttt{.keras}, \texttt{.h5}), which is loaded through \ac{ML} frameworks rather than inspected as source code.

\mypar{Model Persistence and Loading}
Many popular frameworks support persistence mechanisms that range from weights-only storage (largely data-oriented) to full-object serialization that may reconstruct executable objects at load time. For example, PyTorch documents multiple ways to save and load models and explicitly discusses the security implications of its serialization semantics~\cite{PyTorchSaveLoadRun,PyTorchSerialization}. Keras similarly provides whole-model saving and loading APIs, as well as weights-only workflows, and introduces security-oriented settings such as \texttt{safe\_mode} intended to restrict unsafe deserialization paths~\cite{chollet2015keras,KerasSavingLoading}. TensorFlow supports SavedModel and checkpoint-based mechanisms that restore model graphs and parameters using framework-defined loaders~\cite{abadi2016tensorflow,TFSavedModelGuide,TFCheckpointGuide}. 

\subsection{Risks of Model Sharing}\label{subsec:risks}
Model artifacts are not passive inputs: they are processed by \ac{ML} frameworks, creating an attack surface that spans multiple lifecycle phases. During loading, framework loaders deserialize artifacts, resolve model components, and reconstruct in-memory objects; after loading, phases such as inference and training may invoke framework functionality on attacker-controlled state. Across these phases, malicious behavior can manifest as host-level effects, including file-system access, network communication, process creation, or modification of persistent state. Building on insights from open-source software supply chains~\cite{ohm2020backstabber,ladisa2023sok,duan2021towards}, recent work frames this as an \textit{\ac{ML} model supply chain} risk~\cite{meiklejohn2025machine,digregorio2026onthe}, in which untrusted artifacts are executed through complex framework loading and execution paths.

\subsection{Behavioral Profiling and Host-Based Detection}\label{subsec:profiling}
To clarify how the approach in \S~\ref{sec:reference} can be categorized, we briefly review behavioral profiling and host-based detection. A behavioral profile characterizes the expected runtime behavior of software through its observable interactions with the host system, such as system calls and the file, network, and process operations they generate. Profiling of this kind underlies \ac{HIDS}, which monitor host activity and flag deviations from expected behavior~\cite{forrest1996sense,bridges2019survey}. \ac{HIDS} typically follow misuse-based approaches, which match activity against signatures of known attacks, or anomaly-based approaches, which model normal behavior and treat departures from it as suspicious~\cite{satilmis2024asystematic}. The practical value of anomaly-based detection depends on the \emph{profiling strategy}: when profiles are too heterogeneous, normal behavior varies so widely that benign and malicious activity cannot be separated, which is the classical source of the high false-positive rates reported in operational settings. Through this lens, our approach is a form of anomaly-based behavioral profiling, and several of its components can be cast as a \ac{HIDS}; what distinguishes it is the profiling strategy that we identify and formalize for \ac{ML} model execution, grounded in the intuitions of \S~\ref{sec:found_int}.

\section{Related Work}\label{sec:related_work}
Prior work spans \emph{artifact scanners} that analyze model files prior to execution, \emph{restricted loading} that constrains loading at runtime, empirical analysis of model behavior, and broader analyses of the \ac{ML} model supply chain. In this section, we review these directions.

\mypar{Artifact Analysis and Scanners}
Due to the widespread use of Pickle, many tools focus on detecting unsafe deserialization patterns in Pickle-based artifacts. Trail of Bits’ Fickling~\cite{flickling}, for example, performs program analysis of Pickle files by observing the imports and operations triggered during deserialization and comparing them against manually curated allowlists of trusted \ac{ML} libraries. While effective at identifying unexpected imports, this approach relies on predefined allowlists and remains tightly coupled to Pickle semantics and known library behaviors.

Model-sharing platforms integrate scanners into their upload pipelines. The Hugging Face Hub~\cite{HuggingFaceHub}, for instance, applies automated inspection to uploaded artifacts, including generic malware detection via ClamAV~\cite{ClamAV} and VirusTotal~\cite{VirusTotal}, Pickle-specific analysis to enumerate imported modules~\cite{HuggingFacePickleSecurity}, and third-party tools such as \textsc{ModelScan}~\cite{ModelScanProtectAI} from Protect AI~\cite{ProtectAI} and scanners provided by JFrog~\cite{JFrog}. These static scanners analyze model files before execution and rely on predefined signatures, heuristics, or framework-specific rules to flag suspicious constructs, such as the use of specific serialization mechanisms or model components (e.g., Lambda layers). As with traditional malware scanning, their effectiveness depends on the completeness and maintenance of their rulesets and is inherently limited to the formats and threat patterns they support.

\mypar{Restricted Loading Environments}
\textsc{PickleBall}~\cite{pickleball} analyzes serialized Pickle programs together with the source code of the library that produced the model and derives model-specific policies that constrain the operations permitted during deserialization. These policies are then enforced at load time, restricting the model's behavior rather than merely flagging it. While this approach reduces reliance on fixed signatures and provides stronger guarantees than pure artifact scanning, it remains limited to Pickle-based artifacts and focuses exclusively on the loading phase of the model lifecycle. Similarly, beyond its program analysis capabilities, Fickling also supports runtime loading restrictions by hooking the Pickle library and enforcing security checks.

\mypar{Empirical and Dynamic Analyses}
Casey et al.~\cite{casey2024large} conducted the first large-scale exploit instrumentation study of \ac{ML} supply chain attacks on the Hugging Face Hub. Their analysis systematically examined thousands of publicly available models, identified the serialization formats in use, and empirically tested exploitability by injecting malicious payloads into models serialized with unsafe mechanisms. 
Their approach combines Python-level tracing with operating system–level syscall tracing (via \texttt{strace}) during model loading, enabling the observation of interactions such as file operations, process execution, and network access. However, malicious behavior is identified by searching for the presence of a predefined set of syscalls and Python-level execution primitives (e.g., \texttt{socket}, \texttt{connect}, \texttt{execve}, \texttt{chmod}, as well as \texttt{exec} and \texttt{eval}), which are treated as indicative of compromise attempts. While effective for large-scale measurement, this strategy relies on fixed indicators of suspicious behavior rather than on a structured characterization of expected versus unexpected actions during model execution. Their results show that a majority of models using unsafe serialization formats are exploitable and that existing platform-level scanners fail to flag a substantial fraction of vulnerable or malicious artifacts.
\section{Motivation}
\mypar{Relevance and Timeliness}
The attack surface described in \S~\ref{subsec:risks} poses an increasingly urgent problem.
Recent work~\cite{digregorio2026onthe} has shown that neither framework-level mitigations nor hub-integrated scanners reliably detect malicious model artifacts. This is not merely a limitation of a specific unsafe format: even data-oriented formats can enable arbitrary code execution through framework-level loading logic. 

At the same time, code-based formats remain widespread for sharing \ac{ML} models, despite being intrinsically difficult to secure. Pickle is the most prominent example: code execution during deserialization is part of its semantics, and its documentation explicitly warns against loading untrusted artifacts~\cite{PythonPickleDocs}. Nevertheless, Pickle remains embedded in several \ac{ML} persistence workflows, either directly or indirectly~\cite{PyTorchSerialization}, creating practical exploitation paths and contributing to documented vulnerabilities in model-loading flows~\cite{CVE-2024-3660,CVE-2025-1550,CVE-2025-8747,CVE-2025-9905,CVE-2025-9906,CVE-2025-12058,CVE-2025-49655}.

Moreover, the attack surface extends beyond loading: recent attacks also trigger malicious behavior during inference by abusing legitimate framework APIs~\cite{zhu2025my, CVE-2025-32434}. 

Recent talks at venues such as DEF CON 33~\cite{ParzianDefcon33} and Black Hat USA 2025~\cite{bh2025_torchscript} have further highlighted attacks against ML model formats and frameworks, while model hubs continue to integrate new scanners, such as the VirusTotal integration announced by Hugging Face in October 2025~\cite{carreira2025virustotal}. These developments suggest that model security is an active and evolving arms race between framework hardening, platform scanning, and adversarial abuse.

\mypar{Open Problems}
We identify four main open problems from the approaches reviewed in \emph{Related Work} (\S~\ref{sec:related_work}).

\myemphlist{i}{Limited coverage across formats and lifecycle phases} Existing defenses are often tied to a specific artifact format or execution phase. For example, Pickle-oriented approaches such as Fickling~\cite{flickling} and \textsc{PickleBall}~\cite{pickleball} are confined to a single serialization format and primarily act at load time, so they cannot reason about malicious behavior triggered during inference or training. 

\myemphlist{ii}{Trust-shifting mitigations} Some mitigations reduce the amount of executable or framework-resolved state embedded in an artifact, but do not eliminate the underlying trust problem. Weights-only sharing, for instance, reduces loading-time risks, but shifts trust to external components: architectures, custom layers, preprocessing logic, and dependencies must still be distributed separately.

\myemphlist{iii}{Reactive and incomplete scanner coverage} Scanners that generalize across formats, such as those integrated into the Hugging Face Hub~\cite{HuggingFaceHub} (e.g., \textsc{ModelScan}~\cite{ModelScanProtectAI} from Protect AI), typically depend on signatures, heuristics, or framework-specific rules. Their coverage is therefore incomplete and difficult to maintain as frameworks evolve. Recent work systematically bypasses state-of-the-art scanners across alternative loading paths and risky-function gadgets~\cite{liu2025arthideseekmaking}, while prior analysis uncovered Keras vulnerabilities that evaded existing framework mitigations and exposed inconsistencies in scanner support across formats~\cite{digregorio2026onthe}.

\myemphlist{iv}{Unresolved precision--recall tradeoff} Even when detection succeeds, existing approaches still struggle to balance false positives and false negatives. Scanners may conservatively misclassify benign models containing Lambda layers while missing other malicious artifacts~\cite{digregorio2026onthe}, and \textsc{PickleBall} still reports false-positive rates above 20\%~\cite{pickleball}.

\mypar{Motivating Example} A concrete instance of this dynamic is CVE-2025-1550~\cite{CVE-2025-1550}, disclosed in early 2025. Unlike earlier exploits relying on the serialization of executable code (e.g., pickled bytecode or Lambda layers), this vulnerability exploited a previously unexplored execution path in the loading logic of the supposedly safe \texttt{.keras} format, even under Keras \texttt{safe\_mode}~\cite{Digregorio-cve-2025-1550}. By abusing how framework internals resolve and instantiate model components, the attack achieved arbitrary code execution at load time. Following the disclosure, Hugging Face integrated new detection rules from Protect AI that explicitly referenced the CVE~\cite{MorganPAI6Months}, yet other scanners deployed on the Hub still did not flag instances of the same vulnerability weeks later~\cite{digregorio2026onthe}. The episode illustrates the core limitation of static, rule-based scanning: detection is reactive and presupposes prior knowledge of the specific exploitation technique.

\mypar{Toward a Behavior-Based Defense}
The above observations motivate a novel defense that does not rely on a specific file format, serialization mechanism, scanner signature, or known vulnerability. Instead of asking how an attack is encoded inside a model artifact, we focus on the effects this attack, performed via model execution, has on the host system, enforcing execution boundaries (i.e., an execution allowlist) during specific lifecycle phases. We envision this defense as part of practical deployments that can combine static and dynamic analysis. Static analysis can cheaply reject known malicious patterns, while restricted loaders can reduce exposure to specific dangerous mechanisms. Our dynamic approach, instead, is particularly useful for previously unseen attacks, framework abuse, and attacks that manifest outside loading, where static or loader-specific defenses may provide limited coverage.
\section{Threat Model}\label{sec:threat_model}
Expanding the threat model defined by Digregorio et al.~\cite{digregorio2026onthe}, which focuses on arbitrary code execution during the loading phase, we consider the threat posed by malicious \ac{ML} model artifacts that aim to compromise the user’s system during one or more phases of the model lifecycle.

\mypar{Attacker’s Target}
We focus on the execution pipeline composed of the \ac{ML} model's lifecycle phases defined in \S~\ref{subsec:model-lifecycle} (i.e., loading, inference, and training), executed locally on a user’s machine. The system targeted by the attacker consists of: (i) the user environment, including the operating system with an installed \ac{ML} framework; (ii) the model artifact, represented as a serialized file such as \texttt{.keras}~\cite{KerasSerialization} or \texttt{.h5}~\cite{KerasSerialization}; and (iii) the phase-specific framework mechanisms used to operate on the model. Concretely, this includes framework-provided functions such as \texttt{keras.models.load\_model()}~\cite{TFKerasLoadModel} during the loading phase and \texttt{keras.Model.predict()}~\cite{KerasModelTrainingAPIs} during inference.
We focus on scenarios in which a user executes a model obtained from an external source.

\mypar{Attacker’s Goal}
We consider an attacker who crafts a malicious model artifact to compromise the victim’s system during one of the model lifecycle phases. The attacker’s primary objective is to induce attacker-controlled behavior that escapes the execution context of the model and results in control over the host system, enabling arbitrary interactions with system resources, including but not limited to arbitrary code execution, within the privileges of the user executing the model. 
These attacks exploit vulnerabilities in the deserialization process or in the \ac{ML} framework and target the host system, rather than the model’s predictions, as done by \emph{model-level attacks} (e.g., poisoning attacks)~\cite{biggio2018}.

\mypar{Attacker’s Capabilities and Assumptions}
The attacker can create, modify, and distribute \ac{ML} model artifacts but has no prior access to the target’s system. We consider two distribution channels: (1) \emph{public repository poisoning}, in which a malicious model is uploaded to a trusted platform (e.g., Hugging Face Hub~\cite{HuggingFaceHub}, Kaggle~\cite{KaggleModels}, or GitHub) under the guise of a legitimate resource; and (2) \emph{direct delivery}, in which the artifact is sent to the victim through private channels such as email or cloud storage. We do not make specific assumptions about whether the victim enables framework-level mitigations such as \texttt{safe\_mode}~\cite{KerasSavingLoading} in Keras or \texttt{weights\_only}~\cite{PyTorchSaveLoadRun} in PyTorch.
Prior work~\cite{digregorio2026onthe} has shown that such mechanisms are not always sufficient to prevent attacks at model loading time and, in some cases, may create a false sense of security.

While we assume the \ac{ML} framework itself to be trustworthy, the threat model allows an attacker to abuse both model-embedded code and legitimate framework code paths. In particular, an adversary may embed malicious routines within a model artifact or reuse existing framework functionality in unintended ways to achieve malicious effects, as demonstrated by recent vulnerabilities such as CVE-2025-9905~\cite{CVE-2025-9905}, CVE-2025-9906~\cite{CVE-2025-9906}, and CVE-2025-8747~\cite{CVE-2025-8747}.

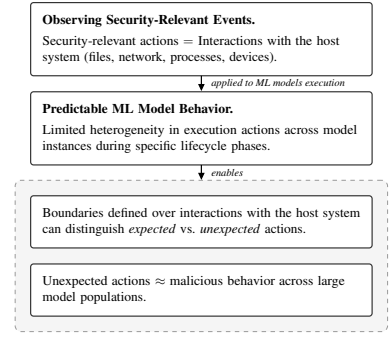
\begin{figure}[t]
    \centering
    \resizebox{0.6\linewidth}{!}{\definecolor{papergrey}{RGB}{245,245,245}
\definecolor{reqred}{RGB}{245,240,240}

\begin{tikzpicture}[
    scale=0.50,
    node distance=0.35cm and 0.9cm,
    base/.style={
        rectangle, rounded corners=2pt, draw=black!80, thick,
        text width=7.5cm, align=left, font=\small, inner sep=8pt
    },
    concept/.style={base, fill=white},
    implementation/.style={base, fill=white},
    requirement/.style={base, fill=reqred},
    arrow/.style={-{Triangle[width=1.5mm]}, thick, black, cap=rect},
    labelstyle/.style={midway, right, xshift=3pt, font=\scriptsize\itshape}
]

    % --- Nodes ---
    
    % 1. Action Model
    \node [concept] (action) {
        \mypar{Observing Security-Relevant Events}\\
        Security-relevant actions $=$ Interactions with the host system (files, network, processes, devices).
    };

    % 2. Predictable Behavior
    \node [concept, below=of action] (predictable) {
        \mypar{Predictable ML Model Behavior}\\
        Limited heterogeneity in execution actions across model instances during specific lifecycle phases.
    };

    % 3. Lifecycle Boundaries
    \node [implementation, below=0.7cm of predictable] (lifecycle) {
        Boundaries defined over interactions with the host system can distinguish \textit{expected} vs. \textit{unexpected} actions.
    };

    % 4. Approximation
    \node [implementation, below=of lifecycle] (approx) {
        Unexpected actions $\approx$ malicious behavior across large model populations.
    };

    % --- Background logical grouping (Superblock) ---
    \begin{scope}[on background layer]
        \node[fill=papergrey, draw=black!40, dashed, rounded corners, inner sep=10pt, fit=(lifecycle) (approx)] (implications) {};
    \end{scope}

    % --- Connections ---
    
    \draw [arrow] (action) -- (predictable) 
        node[labelstyle] {applied to ML models execution};
    
    \draw [arrow] (predictable) -- (implications.north) 
        node[labelstyle] {enables};

\end{tikzpicture}}
    \caption{
    Overview of the core intuitions behind lifecycle-aware execution boundaries for \ac{ML} models.}
    \label{fig:fig_intuitions_summary}
\end{figure}

\section{Foundational Intuitions}\label{sec:found_int}
This section provides an overview of the fundamental intuitions underlying this work, which form its conceptual core. Here, we introduce the notions used throughout the paper, such as software execution actions, and formalize the distinction between \emph{expected}, \emph{unexpected}, and \emph{malicious} actions. 
We then observe that, by leveraging the structured lifecycle of \ac{ML} models, execution boundaries can be instantiated in a precise and enforceable manner across different execution phases. A graphical summary of the conceptual flow underlying this work is shown in Figure~\ref{fig:fig_intuitions_summary}.

\subsection{Observing Security-Relevant Events}\label{subsec:security_events}

We adopt a dynamic approach to secure \ac{ML} model execution. Unlike existing static approaches, we shift the focus away from attack-specific reasoning and instead define an allowlist-based approach centered on \emph{security-relevant events} triggered during model execution.
These events are captured at the boundary between the \ac{ML} execution environment and the host system, rather than through internal framework states or model-specific abstractions. We focus on events that directly impact host security, such as file-system access, network communication, process creation, and device interaction.

An alternative approach would be to instrument the \ac{ML} framework, for example, by monitoring internal APIs or execution paths during model loading, inference, or training. Under our threat model, however, this strategy is unreliable. Even assuming the framework to be trustworthy, attackers can exploit legitimate framework code paths to induce unintended effects~\cite{zhu2025my}, thereby bypassing in-framework monitoring.
Recent vulnerabilities illustrate this limitation. For instance, CVE-2025-9906~\cite{CVE-2025-9906} demonstrates how legitimate Keras code paths can be abused during model loading to disable the \texttt{safe\_mode} security mechanism.
Any observation or enforcement mechanism implemented purely within the framework would be exposed to the same manipulation.

\subsection{Execution Actions}\label{sec:execution_actions}
We define an \emph{action} as any operation executed by a software during its runtime that results in an interaction with the host system on which it runs (e.g., file-system access, network communication, process creation, or memory management). We denote by $\mathcal{A}$ the universe of such actions.% that can be performed on a system.

The classification of actions as expected or unexpected is inherently contextual. We therefore define an \emph{expected action} as an action that is consistent with the intended behavior of the software under a given operational context, as determined by its specification, typical usage, or an explicit security policy defined by a system administrator. 
Formally, we denote by $\mathcal{A}_{\text{exp}} \subseteq \mathcal{A}$ the set of actions expected under a given context, and by $\mathcal{A}_{\text{unexp}} = \mathcal{A} \setminus \mathcal{A}_{\text{exp}}$ the set of unexpected actions. 

For example, it might be considered expected for a text editor to perform a write operation on the path of the edited file (the executed action, which includes information about the accessed path) during the saving phase (a lifecycle phase of the text editor), provided that the system policy allows such file manipulation (as defined by a system administrator). Concretely, writing to a user file such as \texttt{/home/user/note.txt} (opened as part of the edit) would be considered expected. In contrast, a write operation targeting an unrelated file path (e.g., \texttt{/etc/passwd}) would be considered unexpected.

According to the threat model defined in \S~\ref{sec:threat_model}, we further define the subset of \emph{malicious actions} $\mathcal{A}_{\text{mal}} \subseteq \mathcal{A}_{\text{unexp}}$ as those actions whose goal is to harm the system.

\subsubsection{Action Provenance in ML Model Execution}\label{sec:provenance}
In the \ac{ML} model execution domain, a significant portion of the executed actions is determined by code from the \ac{ML} framework. Indeed, each lifecycle phase of an \ac{ML} model is realized through calls to framework APIs, which execute within the framework’s runtime environment while operating on a specific model instance (e.g., during loading, inference, or training). However, additional actions may also originate from code specific to the model instance itself, such as custom model definitions or auxiliary logic serialized within the model artifact. As a result, both framework code and model-embedded code contribute to the observed execution behavior.

Moreover, as clarified in the threat model in \S~\ref{sec:threat_model}, an attacker may abuse both model-embedded code and legitimate framework code paths to achieve malicious effects. Since both execution sources can ultimately lead to arbitrary code execution~\cite{CVE-2025-8747,CVE-2025-9906,CVE-2025-12058}, we treat them uniformly and do not attempt to distinguish their provenance.

For brevity and readability, throughout this section and the rest of the paper, we refer to execution actions as being ``performed by an \ac{ML} model'' or, interchangeably, by ``an \ac{ML} framework.''

\subsubsection{Action Granularity}\label{subsec:granularity}
To capture their security-relevant semantics, actions are identified not solely by their type (e.g., \emph{file read} or \emph{file write}) but also by contextual attributes that determine their effect on the system.
For example, reading \texttt{model.pt} and reading \texttt{/etc/passwd} constitute distinct actions, even though both are file-read operations. Similarly, accessing \texttt{model.pt} in read versus write mode represents different actions.

\subsection{Predictable ML Model Behavior}\label{subsec:fundamental_obs}
Generic software is characterized by high heterogeneity across instances, as distinct programs may legitimately perform radically different actions. A \emph{software instance} denotes a specific realization of a software program, belonging to one or more categories providing similar functionality (e.g., \texttt{nano} is an instance of the text editor category). This variability introduces two challenges: (i) distinguishing expected from unexpected actions is difficult without detailed knowledge of the specific instance; and (ii) the boundary around malicious actions lacks clear delineation. An action cannot be classified as malicious \emph{per se}, but only relative to the execution characteristics and intended functionality of the specific instance: a text editor may legitimately access certain cache paths that another would not, even within the same category.
As a consequence, static allow/deny policies alone are insufficient for malware detection and are better suited to expressing and auditing administrator security policies. This challenge has motivated a broad line of malware-analysis research, from signature-based detection to \ac{ML}-based approaches that distinguish benign from malicious behaviors~\cite{gibert2020rise,pendlebury2019tesseract,aslan2020comprehensive}.

In contrast, \ac{ML} exhibits limited heterogeneity across instances: a model artifact is not a distinct software instance but a variation within the capability space exposed by its \ac{ML} framework, a general property we confirm across frameworks (\S~\ref{sec:evaluation}). This determines our profiling strategy that governs anomaly-based detection (\S~\ref{subsec:profiling}): observe the behavior of a framework that is executing an \ac{ML} model within a certain lifecycle phase. Because behaviors within framework--phase pairs are narrow and stable, expected behaviors (profiles) can be expressed as explicit allowlists, and any action outside them can be treated as anomalous. The low heterogeneity of these profiles is what allows such allowlists to separate benign from malicious behavior, rather than incurring the false-positive rates that burden anomaly-based detection when the profiles are too heterogeneous~\cite{satilmis2024asystematic}. 

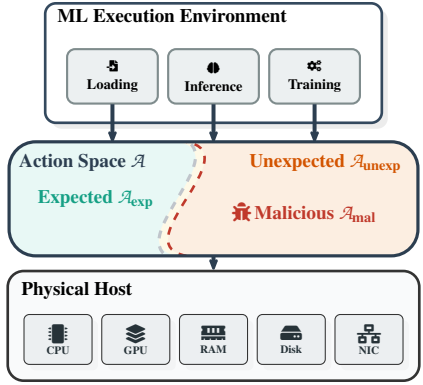
\begin{figure}[t]
    \centering
    % --- Professional Color Palette ---
\definecolor{envFill}{HTML}{FFFFFF}
\definecolor{envHeader}{HTML}{2C3E50}
\definecolor{phaseFill}{HTML}{ECF0F1}
\definecolor{phaseBorder}{HTML}{7F8C8D}
\definecolor{envBorder}{HTML}{34495E}
\definecolor{safeFill}{HTML}{E8F8F5}
\definecolor{safeStroke}{HTML}{16A085}
\definecolor{unsafeFill}{HTML}{FEF9E7}
\definecolor{unsafeStroke}{HTML}{D35400}
\definecolor{malFill}{HTML}{FADBD8}
\definecolor{malStroke}{HTML}{C0392B}
\definecolor{sysFill}{HTML}{F4F6F7}
\definecolor{sysBorder}{HTML}{2C3E50}
% Hardware Colors (From Runtime Image)
\definecolor{hwFill}{HTML}{ECEFF1}
\definecolor{hwStroke}{HTML}{607D8B}

\begin{tikzpicture}[
  font=\sffamily,
  transform shape,
  scale=0.6,
  node distance=0.6cm and 0.2cm,
  % ---------------------
  phase/.style={
    rectangle, rounded corners=3pt, draw=phaseBorder, fill=phaseFill,
    line width=0.8pt, minimum width=2.0cm, minimum height=1.2cm,
    align=center, font=\bfseries\scriptsize,
  },
  % Hardware Node Style
  hwNode/.style={
    rectangle, rounded corners=2pt, fill=hwFill, draw=black!80,
    minimum width=1.5cm, minimum height=1.1cm,
    font=\bfseries\scriptsize\color{hwStroke!40!black},
    align=center, drop shadow={opacity=0.05}
  },
  % Flow Style (Thick, Rounded, Correct Color)
  flow/.style={
    ->, >={Stealth[round, length=6pt]},
    line width=1.5pt,
    color=sysBorder!100,
    rounded corners=4pt
  }
]

% ==========================================================
% CENTRAL AXIS STRATEGY
% ==========================================================
\node[minimum width=8.8cm, minimum height=2.5cm] (ActionRef) at (0,0) {};
\coordinate (EnvCenter) at (0, 2.7);
\coordinate (HostCenter) at (0, -3.1);

% ==========================================
% DRAW: Action Space A
% ==========================================
\begin{scope}
  \path[clip, rounded corners=10pt]
    ($(ActionRef.south west) + (-0.1, 0)$)
    rectangle
    ($(ActionRef.north east) + (0.1, 0)$);

  % Background
  \fill[unsafeFill]
    ($(ActionRef.south west) + (-1, -1)$)
    rectangle
    ($(ActionRef.north east) + (1, 1)$);

  % Expected (Green)
  \fill[safeFill]
    ($(ActionRef.south) + (-0.8, -3)$)
    -- ($(ActionRef.south) + (-0.8, 0)$)
    .. controls ($(ActionRef.center)+(-2.2,-1.0)$) and ($(ActionRef.center)+(0.6,1.0)$) ..
    ($(ActionRef.north) + (-0.8, 0)$)
    -- ($(ActionRef.north) + (-0.8, 3)$)
    -- ($(ActionRef.north west) + (-3, 0)$)
    -- ($(ActionRef.south west) + (-3, 0)$)
    -- cycle;

  % Dividing Line
  \draw[sysBorder!30, line width=1.2pt, dashed]
    ($(ActionRef.south) + (-0.8, -0.1)$)
    .. controls ($(ActionRef.center)+(-2.2,-1.0)$) and ($(ActionRef.center)+(0.6,1.0)$) ..
    ($(ActionRef.north) + (-0.8, 0.1)$);
\end{scope}

% ==========================================
% Malicious Subset — fills the unexpected region with a tiny inset,
% only the S-curve side gets a dashed border (top/right/bottom
% coincide with the action space border itself)
% ==========================================
\begin{scope}
  \path[clip, rounded corners=10pt]
    ($(ActionRef.south west) + (-0.1, 0)$)
    rectangle
    ($(ActionRef.north east) + (0.1, 0)$);

  % Malicious fill: from the S-curve (shifted +0.2) to the right/top/bottom edges
  \fill[malFill, fill opacity=0.35]
    ($(ActionRef.south) + (-0.6, 0)$)
    .. controls ($(ActionRef.center)+(-2.0,-1.0)$) and ($(ActionRef.center)+(0.8,1.0)$) ..
    ($(ActionRef.north) + (-0.6, 0)$)
    -- ($(ActionRef.north east) + (0.2, 0)$)
    -- ($(ActionRef.south east) + (0.2, 0)$)
    -- cycle;

  % Only draw the dashed S-curve border (left side of malicious region)
  \draw[malStroke, line width=1pt, dashed]
    ($(ActionRef.south) + (-0.6, 0)$)
    .. controls ($(ActionRef.center)+(-2.0,-1.0)$) and ($(ActionRef.center)+(0.8,1.0)$) ..
    ($(ActionRef.north) + (-0.6, 0)$);
\end{scope}

% Malicious label
\node[font=\bfseries\large, color=malStroke] at ($(ActionRef.center) + (2.0, -0.3)$) {%
  \faBug\,Malicious $\mathcal{A}_{\text{mal}}$
};

% Labels
\node[font=\bfseries\large\color{unsafeStroke}, anchor=north east, xshift=-4pt, yshift=-4pt]
  at (ActionRef.north east) {Unexpected $\mathcal{A}_{\text{unexp}}$};

\node[anchor=north west, font=\bfseries\large\color{sysBorder}, xshift=0pt, yshift=-4pt]
  at (ActionRef.north west) {Action Space $\mathcal{A}$};

\node[font=\bfseries\large\color{safeStroke}]
  at ($(ActionRef.center) + (-2.6, 0)$) {Expected $\mathcal{A}_{\text{exp}}$};

% Border
\draw[sysBorder, line width=1.2pt, rounded corners=10pt]
  ($(ActionRef.south west) + (-0.1, 0)$) rectangle ($(ActionRef.north east) + (0.1, 0)$);

% ==========================================
% DRAW: ML Model Execution Environment (Unchanged)
% ==========================================
\node[phase] (Infer) at (EnvCenter) {
  \faBrain\\[4pt]
  \small Inference
};
\node[phase, left=0.2cm of Infer] (Load) {
  \faFileImport \\[4pt]
  \small Loading
};
\node[phase, right=0.2cm of Infer] (Train) {
  \faCogs\\[4pt]\small Training
};

% Container
\begin{scope}[on background layer]
\node[
  draw=envBorder, fill=envFill, line width=1pt, rounded corners=4pt,
  fit=(Load)(Train)(Infer),
  inner sep=8pt, inner ysep=12pt, yshift=5pt,
  drop shadow={opacity=0.15}
] (EnvBox) {};
\node[anchor=north west, font=\bfseries\large\color{black}, xshift=5pt, yshift=-1pt]
  at (EnvBox.north west) {ML Execution Environment};
\end{scope}

% ==========================================
% DRAW: Physical Host
% ==========================================
\node[hwNode] (RAM) at (HostCenter) {\large\faMemory\\ RAM};
\node[hwNode, left=0.2cm of RAM] (GPU) {\large\faLayerGroup\\ GPU};
\node[hwNode, left=0.2cm of GPU] (CPU) {\large\faMicrochip\\ CPU};
\node[hwNode, right=0.2cm of RAM] (Disk) {\large\faHdd\\ Disk};
\node[hwNode, right=0.2cm of Disk] (NIC) {\large\faNetworkWired\\ NIC};

\path (RAM.north) ++(0, 0.6cm) coordinate (HostHeader);

\begin{scope}[on background layer]
\node[
  draw=black!80, fill=hwFill!30, line width=1pt, rounded corners=6pt,
  fit=(HostHeader)(CPU)(NIC), inner sep=6pt,
  label={[anchor=north west, font=\bfseries\large\color{black}, xshift=6pt, yshift=-4pt]north west:Physical Host}
] (HostBox) {};
\end{scope}

% ==========================================
% Connections
% ==========================================
\draw[flow] (Load.south) -- (Load.south |- ActionRef.north);
\draw[flow] (Infer.south) -- (Infer.south |- ActionRef.north);
\draw[flow] (Train.south) -- (Train.south |- ActionRef.north);
\draw[flow, dashed] (ActionRef.south) -- (HostBox.north);

\end{tikzpicture}
    \caption{Overview of the action-space abstraction. \ac{ML} lifecycle phases induce actions in a global action space $\mathcal{A}$, which are classified as expected, unexpected, or malicious and manifest as interactions with the host system.}
    \label{fig:fig_abstraction_layer}
\end{figure}

\mypar{Execution Boundaries} This intuition enables the definition of clear \emph{execution boundaries} for \ac{ML} model execution. The boundary between expected and unexpected actions during a given lifecycle phase can be defined once and reused across a large number of model instances. Moreover, the limited heterogeneity makes it feasible to draw this boundary so that $\mathcal{A}_{\text{unexp}}$ is a \emph{minimal} superset of the malicious set defined in \S~\ref{sec:execution_actions}, closely approximating it: an action deemed unexpected across a large model population renders any model performing it suspicious. For example, a model that connects to a remote server when loaded from a local file can be considered suspicious. Figure~\ref{fig:fig_abstraction_layer} provides a conceptual representation of this intuition.

\section{Approach and Implementation}\label{sec:reference}
In this section, we describe how we translate the foundational intuitions introduced in \S~\ref{sec:found_int} into a systematic approach. We first outline the reference design (\sys) that a system based on these intuitions should follow. We then present our reference implementation (\ri), with its technical details, which realizes this design and is used to experimentally validate our intuitions.

\subsection{Reference Design (\sys)}\label{subsec:ref_design}

Our reference design conceptually includes four core components:
\circnum{1} abstraction of low-level events into higher-level actions, \circnum{2} definition of execution boundaries, \circnum{3} orchestration and isolation of \ac{ML} model lifecycle phases, and \circnum{4} tracing and enforcement of execution boundaries.

\circsubsubsection{Action Abstraction}\label{subsec:action_abstraction}
This component enables reasoning about software behavior in terms of semantically meaningful actions with the host system, abstracting low-level execution events (e.g., system calls) into high-level events that collectively instantiate the action space $\mathcal{A}$.
This abstraction is necessary because the same low-level event may correspond to different actions depending on its context, and conversely, the same action may arise from different low-level events. 
Crucially, high-level actions must be defined with sufficient granularity to preserve security-relevant context (see \S~\ref{subsec:granularity}, \textit{Action Granularity}).

\circsubsubsection{Execution Boundaries Definition}\label{sec:execution_boundary}
For each framework--phase pair, our reference design specifies the set of actions (i.e., interactions with the host system) that a model is expected to perform (i.e., $\mathcal{A}_{\text{exp}}$ as in \S~\ref{sec:execution_actions}, \textit{Execution Actions}). This set constitutes the execution boundaries for that pair. 
Execution boundaries may incorporate both actions that are generally expected based on standard practice and domain knowledge (which are typically shared across different frameworks) and actions that depend on framework-specific behavior.

The latter may be derived from empirical observation of legitimate model executions. This empirical derivation, partially supported by expert knowledge, is our primary method and proceeds in three steps. First, a representative set of benign models is executed under the target lifecycle phase, and the observed low-level events are mapped into actions using the \circnum{1}~\emph{Action Abstraction}. Second, correlated actions are aggregated into the same candidate boundary entry; for example, multiple file accesses within the same directory can be represented as a single directory-level entry. Finally, an analyst manually validates each candidate entry.

The manual validation step serves two purposes. It confirms that a candidate boundary entry is legitimate under the assumption that the model artifact is untrusted: the analyst verifies that the entry cannot be abused in its current form; otherwise, the corresponding actions must be aggregated differently or represented at a finer granularity. Following the previous example, this may require one entry for each accessed file rather than a single entry allowing access to the entire directory. Validation also distinguishes genuine model-phase behavior from lazy-initialization effects introduced by libraries or runtime environments; when such effects are legitimate but unrelated to the phase under analysis, they can be excluded by letting the orchestrator handle the corresponding initialization before monitoring begins.

Once execution boundaries are defined for a given lifecycle phase, any action performed during that phase that falls outside the expected set (i.e., belongs to $\mathcal{A}_{\text{unexp}}$) is treated as \emph{suspect}. Depending on the deployment context, such violations may trigger an alert or halt the model execution.

While this process could potentially be further automated, we consider such automation an engineering effort and therefore out of scope for this work.

\circsubsubsection{Orchestrator}\label{subsec:isolating}
In the proposed design, the orchestrator controls \ac{ML} model execution and isolates the lifecycle phase under analysis, ensuring that observed behavior is interpreted within the correct semantic context, as the same model may interact with the host system in different ways across lifecycle phases.
In addition, phase isolation reduces noise introduced by framework routines and initialization.

\circsubsubsection{Tracer}\label{sec:des_tracer}
This component enables the observation of the interactions between the executing \ac{ML} model and the host system. 
To do this, multiple tracing mechanisms can be used (see Appendix~\ref{app:tracing}), each offering different trade-offs in terms of runtime overhead, deployability, and engineering effort. The foundational intuitions are largely orthogonal to the concrete monitoring mechanism employed. 

\subsection{Reference Implementation (\ri)}\label{sec:ref_imp}

\mypar{Implementation Choices}
While \sys describes a general approach that can be applied across different systems, the concrete implementation of the \circnum{1}~\emph{Action Abstraction} and the \circnum{4}~\emph{Tracer} inherently depends on the underlying architecture and operating system. Our reference implementation, \ri, targets Linux-based operating systems on x86-64. This choice is motivated by the widespread adoption of Linux in \ac{ML} environments and by the fact that major \ac{ML} frameworks provide first-class support for Linux platforms (e.g., after version~2.10, TensorFlow no longer provides native GPU support on Windows~\cite{tf_install_pip}).

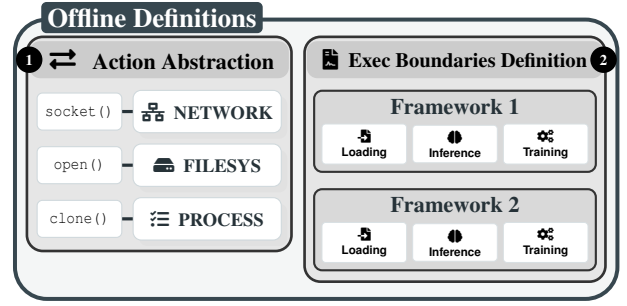
\begin{figure}[t]
    \centering
    \definecolor{mainStroke}{HTML}{37474F}
\definecolor{mainFill}{HTML}{F4F6F6}
\definecolor{lineColor}{HTML}{263238}
\definecolor{subLine}{HTML}{CFD8DC}
\definecolor{bgCard}{HTML}{FFFFFF}
\definecolor{orchFill}{HTML}{E8EAF6}
\definecolor{orchStroke}{HTML}{3949AB}
\definecolor{orchHeader}{HTML}{C5CAE9}
\definecolor{traceFill}{HTML}{E3F2FD}
\definecolor{traceStroke}{HTML}{1976D2}
\definecolor{traceHeader}{HTML}{BBDEFB}
\definecolor{commonFill}{HTML}{E0F2F1}
\definecolor{specFill}{HTML}{E3F2FD}
\definecolor{safeText}{HTML}{00695C}
\definecolor{traceText}{HTML}{1565C0}
\definecolor{codeBg}{HTML}{ECEFF1}
\definecolor{codeText}{HTML}{455A64}

\begin{tikzpicture}[
    font=\sffamily,
    transform shape,
    scale=0.7, 
    node distance=0.4cm,
    % 1. Header Styles
    compHeader/.style={
        rectangle, rounded corners=4pt, 
        minimum height=0.7cm, 
        minimum width=4.9cm,
        font=\bfseries\large,
        align=center
    },
    compHeaderb/.style={
        rectangle, rounded corners=4pt, 
        minimum height=0.7cm, 
        minimum width=5.6cm,
        font=\bfseries\large,
        align=center
    },
    % 2. Section Body (Fixed width for alignment)
    sectionBody/.style={
        rectangle, rounded corners=6pt, line width=1pt,
        minimum width=3.5cm, 
        inner ysep=0pt,
        align=center
    },
    badge/.style={
        circle, fill=mainStroke, text=white, font=\bfseries\footnotesize,
        minimum size=0.5cm, inner sep=0pt, anchor=center
    },
    policyCard/.style={
        rectangle, rounded corners=3pt, draw=subLine,
        line width=0.8pt, 
        minimum width=4cm, 
        align=center, 
        font=\bfseries\small\color{lineColor},
        inner sep=4pt
    },
    % Phase Box Style
    phaseBox/.style={
        rectangle, rounded corners=2pt,
        draw=subLine!80, fill=bgCard,
        minimum width=1.6cm, inner sep=3pt,
        font=\scriptsize\sffamily\color{black}, align=center,
    },
    codeBox/.style={
        rectangle, rounded corners=2pt, fill=white, draw=subLine,
        minimum width=1.6cm, minimum height=0.7cm,
        font=\ttfamily\small\color{codeText}, align=center
    },
    resourceBox/.style={
        rectangle, rounded corners=3pt, fill=bgCard, draw=subLine,
        minimum width=2.8cm, minimum height=0.8cm, 
        font=\bfseries\normalsize\color{lineColor}, align=center,
        drop shadow={opacity=0.05}
    },
    arrowStyle/.style={
        -, line width=1.2pt, color=lineColor
    }
]

% --- LAYER SETUP ---
\pgfdeclarelayer{background}
\pgfdeclarelayer{middle}
\pgfsetlayers{background,middle,main}

% ==========================================================
% CONTENT ELEMENTS
% ==========================================================

% --- Group 1: Syscall-to-Interaction (Left) ---
\node[codeBox] (Sock) at (0, 6.5) {\color{black}socket()};
\node[resourceBox, right=0.2cm of Sock] (NetRes) {\faNetworkWired\ \ NETWORK};
\draw[arrowStyle] (Sock.east) -- (NetRes.west);

\node[codeBox, below=0.3cm of Sock] (Open) {\color{black}open()};
\node[resourceBox, at=(Open-|NetRes)] (FSRes) {\faHdd\ \ FILESYS};
\draw[arrowStyle] (Open.east) -- (FSRes.west);

\node[codeBox, below=0.3cm of Open] (Clone) {\color{black}clone()};
\node[resourceBox, at=(Clone-|NetRes)] (ProcRes) {\faTasks\ \ PROCESS};
\draw[arrowStyle] (Clone.east) -- (ProcRes.west);

% --- Group 2: Boundary Definitions (Right) ---
\node[policyCard, fill=black!15, draw=black!80, right=0.6cm of NetRes.north east, anchor=north west] (Base) {
    {\large Framework 1} \\[0.15cm]
    \begin{tikzpicture}[node distance=0.1cm, transform shape=false]
        \node[phaseBox] (p1) {\faFileImport\\ Loading};
        \node[phaseBox, right=of p1] (p2) {\faBrain\\ Inference};
        \node[phaseBox, right=of p2] (p3) {\faCogs\\ Training};
    \end{tikzpicture}
};

\node[policyCard, fill=black!15, draw=black!80, below=0.3cm of Base] (Framework) {
    {\large Framework 2} \\[0.15cm]
    \begin{tikzpicture}[node distance=0.1cm, transform shape=false]
        \node[phaseBox] (fp1) {\faFileImport\\ Loading};
        \node[phaseBox, right=of fp1] (fp2) {\faBrain\\ Inference};
        \node[phaseBox, right=of fp2] (fp3) {\faCogs\\ Training};
    \end{tikzpicture}
};

% ==========================================================
% BACKGROUND BOXES (MIDDLE LAYER)
% ==========================================================
\begin{pgfonlayer}{middle}
    % Blue Box (Syscalls)
    \node[sectionBody, fill=black!10, draw=black!80, fit=(Sock) (NetRes) (ProcRes), inner ysep =12pt, yshift=8pt] (MapBody) {};

    % Purple Box (Boundaries)
    \node[sectionBody, fill=black!10, draw=black!80, fit=(Base) (Framework), inner ysep=12pt, yshift=8pt ] (BoundBody) {};
\end{pgfonlayer}

% ==========================================================
% MAIN CONTAINER (BACKGROUND LAYER)
% ==========================================================
\begin{pgfonlayer}{background}
    \node[rectangle, rounded corners=12pt, line width=1.5pt, draw=mainStroke, fill=mainFill,
        fit=(BoundBody) (MapBody), inner sep=4pt, inner ysep=6pt] (MainBox) {};
\end{pgfonlayer}

% ==========================================================
% HEADERS & TITLES (MAIN LAYER)
% ==========================================================
\node[fill=mainStroke, text=white, font=\bfseries, rounded corners=2pt, anchor=west, xshift=15pt] 
    at (MainBox.north west) {\Large Offline Definitions};

\node[compHeader, fill=black!20, anchor=north, yshift=-0.1cm] (MapHeader) at (MapBody.north) 
    {{\fontsize{14pt}{14pt} \color{black}\faExchange* \ \ Action Abstraction}};
\node[badge, fill=black] at (MapHeader.west) {1};

\node[compHeaderb, fill=black!20, anchor=north, yshift=-0.1cm] (BoundHeader) at (BoundBody.north) 
    {{\fontsize{11pt}{11pt}\selectfont\color{black}\faFileContract\ \ Exec Boundaries Definition}};
\node[badge, fill=black] at (BoundHeader.east) {2};

\end{tikzpicture}
    \caption{Offline definitions of \ri, showing the \textit{Action Abstraction} and the \textit{Execution Boundaries Definition}.}
    \label{fig:offline_config}
\end{figure}

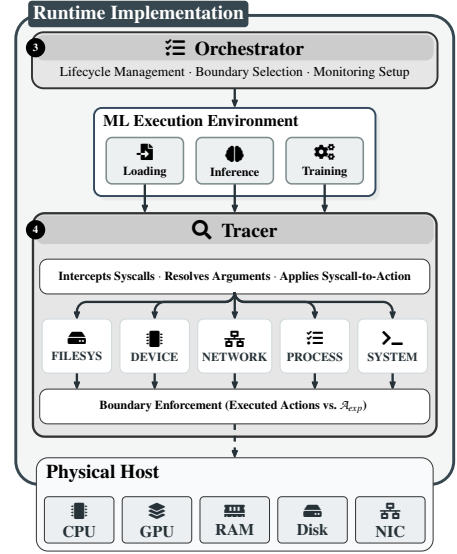
\begin{figure}[t] 
    \centering 
    \resizebox{0.7\linewidth}{!}{ 
        % --- Color Palette ---
\definecolor{mainStroke}{HTML}{37474F}
\definecolor{mainFill}{HTML}{F4F6F6}
\definecolor{lineColor}{HTML}{263238}
\definecolor{subLine}{HTML}{CFD8DC}
\definecolor{bgCard}{HTML}{FFFFFF}

% Component Colors
\definecolor{orchFill}{HTML}{E8EAF6}
\definecolor{orchStroke}{HTML}{3949AB}
\definecolor{orchHeader}{HTML}{C5CAE9}

\definecolor{envFill}{HTML}{FFFFFF}
\definecolor{envHeader}{HTML}{2C3E50}
\definecolor{envBorder}{HTML}{34495E}
\definecolor{phaseFill}{HTML}{ECF0F1}
\definecolor{phaseBorder}{HTML}{7F8C8D}

\definecolor{safeFill}{HTML}{E8F8F5}
\definecolor{safeStroke}{HTML}{009688}
\definecolor{unsafeFill}{HTML}{FFF8E1}
\definecolor{unsafeStroke}{HTML}{FF8F00}
\definecolor{malFill}{HTML}{FFEBEE}
\definecolor{malStroke}{HTML}{C62828}
\definecolor{sysBorder}{HTML}{455A64}

\definecolor{traceFill}{HTML}{E3F2FD} 
\definecolor{traceStroke}{HTML}{1976D2}
\definecolor{traceHeader}{HTML}{BBDEFB} 

\definecolor{enfFill}{HTML}{E8F5E9}
\definecolor{enfStroke}{HTML}{2E7D32}
\definecolor{enfHeaderColor}{HTML}{C8E6C9} 

\definecolor{hwFill}{HTML}{ECEFF1}
\definecolor{hwStroke}{HTML}{607D8B}

\begin{tikzpicture}[
    font=\sffamily,
    transform shape,
    scale=0.7,
    trim left=-1.5cm,
    trim right=3.9cm,
    node distance=0.8cm and 0.4cm,
    % --- Core Styles ---
    compHeader/.style={
        rectangle, rounded corners=4pt, 
        minimum height=0.6cm,     
        minimum width=10.3cm,     
        font=\bfseries\Large,
        align=center
    },
    badge/.style={
        circle, fill=black, text=white, font=\bfseries\footnotesize,
        minimum size=0.5cm, inner sep=0pt, anchor=center
    },
    commonBody/.style={
        rectangle, rounded corners=6pt, line width=1pt,
        minimum width=10.5cm,
        align=center
    },
    orchBody/.style={commonBody, fill=black!10, draw=black!80, minimum height=1.45cm},
    traceBody/.style={commonBody, fill=black!10, draw=black!80, line width=1.2pt, minimum width=5cm, align=center},
    enfBody/.style={commonBody, fill=enfFill, draw=enfStroke},
    % --- Internal Sub-Box Style (Standardized with Blue Text) ---
    subBox/.style={
        rectangle, rounded corners=4pt, draw=black!80, fill=white,
        minimum width=10.1cm, minimum height=0.9cm,
        text=orchStroke!80!black, 
        align=center, font=\small\bfseries
    },
    phase/.style={
        rectangle, rounded corners=3pt, draw=phaseBorder, fill=phaseFill,
        line width=0.8pt, minimum width=2.0cm, minimum height=1.2cm,
        align=center, font=\bfseries\scriptsize,
    },
    classNode/.style={
        rectangle, rounded corners=3pt, fill=bgCard, draw=subLine,
        minimum width=1.8cm, minimum height=1.4cm, 
        align=center, font=\scriptsize\bfseries\color{lineColor}
    },
    hwNode/.style={
        rectangle, rounded corners=2pt, fill=hwFill, draw=black!80,
        minimum width=1.8cm, minimum height=1.2cm,
        font=\bfseries\scriptsize\color{hwStroke!40!black}, align=center,
    },
    flow/.style={
        ->, >={Stealth[round, length=6pt]}, line width=1.2pt,
        color=lineColor, rounded corners=5pt
    }
]

% --- 1. Content Components ---

% Orchestrator
\node[orchBody] (Orch) at (0, 9.1) {};
\node[compHeader, fill=black!20, anchor=north, yshift=-0.08cm] (OrchHeader) at (Orch.north) 
    {\color{black}\faTasks\ \ Orchestrator};
\node[badge] at (OrchHeader.west) {3};
\node[font=\normalsize\color{black}, below=1pt of OrchHeader, align=center] 
    {Lifecycle Management $\cdot$ Boundary Selection $\cdot$ Monitoring Setup};

% ML Execution Environment Components
\node[phase] (Infer) at (0, 6.5) {\large\faBrain\\[4pt] \small Inference};
\node[phase, left=0.3cm of Infer] (Load) {\large\faFileImport\\[4pt] \small Loading};
\node[phase, right=0.3cm of Infer] (Train) {\large\faCogs\\[4pt] \small Training};

% % Action Space
% \node[minimum width=8.5cm, minimum height=1.2cm] (ActionRef) at (0, 6.2) {};
% \begin{scope}
%     \path[clip, rounded corners=8pt] ($(ActionRef.south west) + (-0.1, 0)$) rectangle ($(ActionRef.north east) + (0.1, 0)$);
%     \fill[unsafeFill] ($(ActionRef.south west) + (-1, -1)$) rectangle ($(ActionRef.north east) + (1, 1)$);
%     \fill[safeFill] ($(ActionRef.south) + (-0.8, -3)$) -- ($(ActionRef.south) + (-0.8, 0)$) 
%         .. controls ($(ActionRef.center)+(-2.0,-0.5)$) and ($(ActionRef.center)+(0.6,0.5)$) .. 
%         ($(ActionRef.north) + (-0.8, 0)$) -- ($(ActionRef.north) + (-0.8, 3)$) -- 
%         ($(ActionRef.north west) + (-3, 0)$) -- ($(ActionRef.south west) + (-3, 0)$) -- cycle;
%     \draw[sysBorder!30, line width=1.2pt, dashed] ($(ActionRef.south) + (-0.8, -0.1)$)
%           .. controls ($(ActionRef.center)+(-2.0,-0.5)$) and ($(ActionRef.center)+(0.6,0.5)$) ..
%           ($(ActionRef.north) + (-0.8, 0.1)$);
% \end{scope}
% \node[rectangle, rounded corners=4pt, draw=malStroke, fill=malFill, dashed, line width=1pt, minimum width=4.3cm, minimum height=0.9cm] (MalBox) at ($(ActionRef.center) + (2.0, 0)$) {};
% \node[anchor=north west, font=\bfseries\color{sysBorder}, xshift=5pt, yshift=-1pt] at (ActionRef.north west) {\Large Action Space $\mathcal{A}$};
% \draw[sysBorder, line width=1pt, rounded corners=8pt] ($(ActionRef.south west) + (-0.1, 0)$) rectangle ($(ActionRef.north east) + (0.1, 0)$);

% --- TRACING SHELL INTERNAL COMPONENTS ---
\node[subBox] (TraceText) at (0, 3.5) {\color{black} Intercepts Syscalls $\cdot$ Resolves Arguments $\cdot$ Applies Syscall-to-Action};

% Horizontal row of interaction classes
\node[classNode] (NetMap) at (0, 1.7) {\color{black}\large\faNetworkWired\\[4pt]\small NETWORK};%\tiny (Connect, Bind...)};
\node[classNode, left=0.2cm of NetMap] (DeviceMap) {\color{black}\large\faMicrochip\\[4pt]\small DEVICE};%\tiny (GPU, ...)};
\node[classNode, left=0.2cm of DeviceMap] (FSMap) {\color{black}\large\faHdd\\[4pt]\small FILESYS};%\small (Read, Write...)};
\node[classNode, right=0.2cm of NetMap] (ProcMap) {\color{black}\large\faTasks\\[4pt]\small PROCESS};%\tiny (Fork, IPC...)};
% FIXED: Corrected reference to 'ProcMap' so SYSTEM aligns horizontally
\node[classNode, right=0.2cm of ProcMap] (SysMap) {\color{black}\large\faTerminal\\[4pt]\small SYSTEM};%\tiny (Get Info, Ctrl...)};

\node[subBox, below=0.4cm of NetMap] (EnfBox) {
    {\color{black} Boundary Enforcement (Executed Actions vs. $\mathcal{A}_{exp}$)}
};

% Hardware Components
\coordinate (HWCenter) at (0, -2.8);
\node[hwNode] (RAM) at (HWCenter) {\large\faMemory\\[4pt]\large RAM};
\node[hwNode, left=0.2cm of RAM] (GPU) {\large\faLayerGroup\\[4pt]\large GPU};
\node[hwNode, left=0.2cm of GPU] (CPU) {\large\faMicrochip\\[4pt]\large CPU};
\node[hwNode, right=0.2cm of RAM] (Disk) {\large\faHdd\\[4pt]\large Disk};
\node[hwNode, right=0.2cm of Disk] (NetHW) {\large\faNetworkWired\\[4pt]\large NIC};

% --- 2. Background Layers ---

\begin{pgfonlayer}{background}
    % The main gray container
    \node[rectangle, rounded corners=12pt, draw=mainStroke, fill=mainFill, line width=1.5pt, 
          fit=(Orch)(EnfBox)(Load)(Train), 
          inner xsep=8pt, inner ysep=10pt, minimum height=8.5cm, yshift=-0.4cm] (MainBox) {};
    
    \node[fill=mainStroke, text=white, font=\bfseries\small, rounded corners=2pt, anchor=west, xshift=10pt] 
        at (MainBox.north west) {\Large Runtime Implementation};

    % ML Execution Environment Grouping
    \path (Infer.north) ++(0, 0.45cm) coordinate (EnvHeaderSpacer); 
    \node[draw=envBorder, fill=envFill, line width=1pt, rounded corners=4pt, 
          fit=(EnvHeaderSpacer)(Load)(Train)(Infer), 
          inner sep=6pt, drop shadow={opacity=0.1}] (MLGroup) {};
    \node[anchor=north west, font=\bfseries\color{black}, xshift=4pt, yshift=-2pt] 
        at (MLGroup.north west) {\large ML Execution Environment};

    % The Blue Tracing Shell (Now fits the row correctly)
    \node[traceBody, fit=(TraceText)(FSMap)(SysMap)(EnfBox), inner ysep=15pt, yshift=0.3cm] (TracingContainer) {};
    
    \node[compHeader, fill=black!20, anchor=north, yshift=-0.08cm] (TraceHeader) at (TracingContainer.north) 
        {\color{black}\faSearch\ \ Tracer};
    \node[badge] at (TraceHeader.west) {4};

    % Hardware Frame
    \path (RAM.north) ++(0, 0.8cm) coordinate (HWHeaderSpacer); 
    \node[rectangle, rounded corners=6pt, draw=black!80, fill=hwFill!30, 
          fit=(HWHeaderSpacer)(CPU)(NetHW), 
          inner sep=4pt, 
          label={[anchor=north west, font=\bfseries\small\color{black}, xshift=4pt, yshift=-2pt]north west:\Large Physical Host}] (HWFrame) {};
\end{pgfonlayer}

% --- 3. Flow Connections ---
\draw[flow] (Orch.south) -- (MLGroup.north); 
\draw[flow] (Load.south) -- (Load.south |- TracingContainer.north);
\draw[flow] (Infer.south) -- (Infer.south |- TracingContainer.north);
\draw[flow] (Train.south) -- (Train.south |- TracingContainer.north);

% \draw[flow] (ActionRef.south) -- (TracingContainer.north);
\draw[flow] (TraceText.south) -- ++(0,-0.2) -| (FSMap.north);
\draw[flow] (TraceText.south) -- ++(0,-0.2) -| (NetMap.north);
\draw[flow] (TraceText.south) -- ++(0,-0.2) -| (SysMap.north);
\draw[flow] (TraceText.south) -- ++(0,-0.2) -| (DeviceMap.north);
\draw[flow] (TraceText.south) -- ++(0,-0.2) -| (ProcMap.north);

\draw[flow] (FSMap.south) -- (FSMap.south |- EnfBox.north);
\draw[flow] (NetMap.south) -- (NetMap.south |- EnfBox.north);
\draw[flow] (SysMap.south) -- (SysMap.south |- EnfBox.north);
\draw[flow] (ProcMap.south) -- (ProcMap.south |- EnfBox.north);
\draw[flow] (DeviceMap.south) -- (DeviceMap.south |- EnfBox.north);

\draw[flow, dashed] (EnfBox.south) -- (HWFrame.north);

\end{tikzpicture} 
    }
    \caption{Runtime architecture of \ri, showing the \textit{Orchestrator} and the \textit{Tracer}.} 
    \label{fig:runtime_architecture} 
\end{figure}

\mypar{Architecture}
\ri consists of four main components, referenced using the same naming conventions introduced in the reference design (\S~\ref{subsec:ref_design}): 
\circnum{1}~\emph{Action Abstraction},
\circnum{2}~\emph{Execution Boundaries Definition},
\circnum{3}~\emph{Orchestrator}, and
\circnum{4}~\emph{Tracer}.
The first two components are defined offline prior to execution, whereas the latter two are active at runtime.
Figure~\ref{fig:offline_config} provides a high-level overview of the offline configuration and Figure~\ref{fig:runtime_architecture} shows the runtime
architecture.

\circsubsubsection{Action Abstraction}
The syscall-to-action mapping associates each system call available on a Linux x86-64 system with one or more abstract \emph{interactions with the host system} (i.e., an action in $\mathcal{A}$). 
The mapping explicitly accounts for the fact that the same system call may correspond to different actions depending on its arguments. For example, the \texttt{mmap} system call may represent either anonymous memory allocation or access to a file, depending on whether a file descriptor is provided and on flags such as \texttt{MAP\_ANONYMOUS}. Similarly, \texttt{clone} and \texttt{clone3} may correspond to thread creation or process creation depending on flags such as
\texttt{CLONE\_THREAD}.  

Figure~\ref{fig:categories} (appendices) summarizes the action categories used in \ri, including sub-actions specializing each category. This set is not intended to be exhaustive, nor to represent a canonical taxonomy of system behavior. Rather, it is designed to be sufficiently expressive to capture the security-relevant actions exercised by the \ac{ML} frameworks during specific lifecycle phases. We consider refinements of this set to be possible but orthogonal to the core ideas explored in this work, and dependent on specific use cases.

\circsubsubsection{Execution Boundaries Definition}
We instantiate the boundary-definition process described in \S~\ref{subsec:ref_design} by profiling benign model executions and manually validating distinct action patterns. We first construct baseline boundaries from 600 programmatically generated benign artifacts, comprising 100 models for each of the following categories: PyTorch models relying exclusively on built-in libraries (i.e., fully self-contained Pickle artifacts that also include the model architecture), PyTorch models using custom classes (weights only), Torchvision models (weights only), Keras models in the \texttt{.keras} format (self-contained), Keras models in the legacy \texttt{.h5} format (self-contained), and TensorFlow models in the SavedModel format (self-contained). To cover framework paths that may not be exercised by generated artifacts, we further analyze approximately 1{,}000 models from the Hugging Face Hub.

We execute all these artifacts under \ri across the corresponding lifecycle phases and record the observed actions. For each new candidate entry, we manually verify that the action is legitimate under the assumption that the model artifact is untrusted and identify whether its root cause lies in framework behavior or runtime initialization.

Importantly, boundary construction is performed before evaluating malicious PoCs and in-the-wild models, and no action observed from malicious artifacts is used to define or update the boundaries. A high-level representation of the boundary definitions for each framework--phase pair is shown in appendices (Listings~\ref{lst:high_level_boundary_pytorch_loading},~\ref{lst:high_level_boundary_example},~\ref{lst:high_level_boundary_keras_loading},~\ref{lst:high_level_boundary_keras_inference},~\ref{lst:high_level_boundary_tensorflow_inference}), while the complete definitions are included in the released code.

\circsubsubsection{Orchestrator}
It is responsible for coordinating execution and monitoring, invoking the right framework API (e.g., loading a PyTorch model via \texttt{torch.load}) while enabling tracing. It isolates lifecycle phases by precisely delimiting the execution window to be monitored. Since the code invoking the framework API is controlled by the orchestrator itself, \ri uses process control and synchronization mechanisms to mark the beginning and the end of the lifecycle phase under analysis. 
In particular, two signals are injected immediately before and after the execution of the corresponding framework API, serving as delimiters for the analysis. 
These signals are intercepted by the \emph{Tracer} (\circnum{4}), which starts and stops system-call tracing accordingly.
A high-level example is shown below:
\begin{minted}{python}
...
os.kill(os.getpid(), signal.SIGUSR1)
model = framework.load("path_to_model")
os.kill(os.getpid(), signal.SIGUSR2)
...
\end{minted}
To further limit the noise during the lifecycle phase under analysis, the orchestrator also injects all code required to complete initialization prior to invoking the framework API. This includes framework imports, device initialization (e.g., GPU availability checks), and other setup operations. 

Lastly, this component selects the boundary definition corresponding to the target framework--phase pair from those defined by the \textit{Execution Boundaries Definition} (\circnum{2}).

\circsubsubsection{Tracer}\label{sec:imp_tracer}
The tracing mechanism is built on top of \texttt{libdebug}~\cite{libdebug_2024,libdebug_poster}, an open-source programmatic debugger that exposes ptrace-based debugging functionality through Python APIs, including breakpoint, signal handling, and system-call tracing.
We use its support for user-defined callbacks to install a callback at the entry of each system call executed by the target process, i.e., the process running the \ac{ML} framework during the lifecycle phase under analysis.

Within the callback, the intercepted system call and its arguments are mapped to the corresponding high-level interaction with the host system using the \textit{Action Abstraction} (\circnum{1}). 
The resulting action is then checked against the expected execution boundaries defined through the \textit{Execution Boundaries Definition} (\circnum{2}).
If a boundary violation is detected, a security action is triggered. In \ri, this action consists of logging the violation together with relevant contextual information, such as the system call identifier and its arguments.

\subsubsection{Runtime Overhead}
When measured on the boundary-definition dataset using a laptop with an Intel Core i7-9750H CPU (12 cores, up to 4.5\,GHz) and 64\,GB of RAM, with CPU-only execution for the ML frameworks, setup overhead ranges from $0.126 \pm 0.064$s to $2.273 \pm 0.104$s, depending on the framework--phase pair, and is constant across runs. This setup phase includes the initialization performed by the orchestrator before tracing begins, including the lazy-loading initialization identified during boundary construction. Tracing overhead ranges from $0.001 \pm 0.000$s to $3.224 \pm 0.913$s.

For inference, overhead has low correlation with model size because matrix operations do not generate system calls. For loading, overhead grows approximately linearly with model size, as I/O-related system calls scale with artifact size; however, the relative overhead decreases as the baseline loading time dominates.

These results make \ri suitable as an artifact scanner, which is the scope of our implementation and the focus of our experimental evaluation. We also recognize the potential application of our approach to real-time monitoring. In that setting, the overall approach would remain unchanged, but the syscall tracing mechanism should differ (i.e., the \circnum{4}~\emph{Tracer}): instead of ptrace-based tracing, which incurs frequent context switches, a production-oriented implementation should rely on lower-overhead monitoring mechanisms (Appendix~\ref{app:tracing}).

\subsection{Manual Effort Estimation}
Across the five framework--phase pairs considered in our implementation, namely PyTorch loading and inference, Keras loading and inference, and TensorFlow inference (see \S~\ref{sec:evaluation}), each boundary contains between 4 and 21 entries. Many entries are shared across frameworks or phases, yielding 27 unique entries overall. Lazy-initialization handling is harder to quantify: a single initialization statement may cover multiple lazy-loading effects, while a single library may require multiple initialization steps. Under a conservative accounting, we estimate $\approx 20$ such initializations.

In our experience, validating a previously unseen action requires between 10 minutes and 3 hours for an experienced analyst, depending on the complexity of the framework path that produced it. Overall, an analyst with at least three years of experience in ML and systems security can add support for a new framework from scratch, covering loading and inference, in roughly one to two working days.

Reducing this manual validation effort, for example, through automated profiling, clustering, or root-cause analysis of newly observed benign actions, is left as future work.
\section{Experimental Evaluation}\label{sec:evaluation}
Our experimental evaluation is guided by three research questions, each addressed by a dedicated experiment and corresponding subsection. Specifically, these are:

\begin{marcosboxgreyopen}
\textbf{RQ1.} Is our solution able to identify attacks exploiting novel and previously unseen vulnerabilities?
\end{marcosboxgreyopen}

\begin{marcosboxgreyopen}
\textbf{RQ2.} How does our solution perform against real-world models collected in the wild?
\end{marcosboxgreyopen}

\begin{marcosboxgreyopen}
\textbf{RQ3.} How does our solution compare against state-of-the-art approaches?
\end{marcosboxgreyopen}

\mypar{Loading Settings}
To maximize compatibility, we disable framework-provided security mechanisms (e.g., Keras \texttt{safe\_mode} and PyTorch \texttt{weights\_only}). While reducing the attack surface, these mechanisms can impose practical limitations on loading certain (even benign) model artifacts, which may lead users to disable them. Consequently, this choice ensures that potential threats are not excluded from the analysis and aligns with our threat model.

\mypar{Dependencies Management}
We install, in the containerized environment used for our experiments, all mandatory dependencies required to execute the evaluated models across different lifecycle phases (e.g., the official Keras, PyTorch, TensorFlow, and Torchvision libraries). We then proceed as follows for each experiment:

\myemphlist{i}{Unseen Exploits Simulation}
No additional dependencies beyond the mandatory ones were required, except for TensorFlow I/O~\cite{TensorFlowIO2018}. Dependency management primarily involved selecting the appropriate framework versions to ensure that each vulnerability could be reproduced.

\myemphlist{ii}{Large-scale In-the-Wild Analysis}
We acknowledge that some models may require additional dependencies at runtime. Supporting such dependencies would require the development of an automated dependency-resolution mechanism, which is outside the scope of this work and of the experiment itself. We therefore perform the analysis using only the mandatory dependencies. To ensure transparency and avoid inaccurate results, we implement a mechanism that logs execution information during analysis and excludes models with missing dependencies from classification by \ri. Indeed, missing dependencies may cause exceptions before a model’s actual behavior can be observed, potentially leading to %inaccurate results 
false negatives. As discussed in \S~\ref{sec:largehf}, this limitation does not prevent us from analyzing the vast majority of the identified models (77{,}974 out of 88{,}789), which is largely compatible with the scope of the test (i.e., evaluation over a large number of in-the-wild models). Further discussion on the dependency limitation, as well as possible directions for automating module resolution (which pertain to the reference implementation rather than to the proposed technique), is provided in \S~\ref{sec:limitations}.

\myemphlist{iii}{State-of-the-Art Comparison}
The smaller number of artifacts and the limited set of dependencies compared to large-scale analysis make it possible to identify and manually install all required external dependencies. We also update the orchestrator to account for the import and initialization of these additional libraries. These choices enable a comprehensive and fair comparison with the state of the art across the entire dataset, rather than a reduced subset that could introduce bias. Despite the additional dependencies, no changes are made to the execution boundaries.

\mypar{Scope} Our evaluation does not restrict the analysis to a specific model architecture or family. \ri applies to any artifact supported by the tested frameworks and included in the evaluated datasets or hosted on Hugging Face. This includes artifacts spanning different architectural classes and application domains, from convolutional and fully connected networks to transformer-based language models.

\mypar{Result Validation} We manually validate all results by confirming that models flagged as malicious indeed exhibit malicious behavior. In addition, we verify that the observed execution-boundary violations are consistent with the exploits embedded in the artifacts.

\subsection{Unseen Exploits Simulation (RQ1)}\label{sec:cveabuse}
Our goal is to assess the ability of \ri (and of its underlying intuitions) to detect attacks that exploit novel vulnerabilities in \ac{ML} frameworks.
Relying on \acp{PoC} from real-world CVEs and on the recently proposed TensorAbuse technique by Zhu et al.~\cite{zhu2025my}, which affects popular \ac{ML} frameworks, we simulate a scenario in which a defender must protect against an attacker exploiting such vulnerabilities. We evaluate \sys using framework versions that are vulnerable to these vulnerabilities or by disabling the security mechanisms implemented to mitigate them.
Because our solution does not rely on how an exploit is implemented, but rather on its observable effects on the host system, no vulnerability-specific rules are hard-coded into \ri. As a result, rolling back framework versions approximates the scenario in which such exploits are encountered for the first time, prior to public disclosure and the availability of tailored detection mechanisms. 

This analysis is further complemented by an evaluation against a vulnerability not publicly disclosed at the time of testing, which abuses \texttt{TFSMLayer} in Keras. We identified this vulnerability during the development of \ri while analyzing the behavior of benign models using this layer.
Thanks to the highly informative nature of our approach (which reports the type of violation, accessed paths, and related contextual details), we identified security concerns related to \texttt{TFSMLayer}. In particular, this layer loads external SavedModel artifacts during Keras model loading, which cannot be considered secure~\cite{TFSecurityPolicy}, as demonstrated by TensorAbuse~\cite{zhu2025my}. This behavior effectively bypasses \texttt{safe\_mode} when \texttt{TFSMLayer} is used. We privately reported these observations to the Keras team, who confirmed the issue (already known internally). 
This vulnerability provides a concrete case study that closely mirrors our motivating example, representing a novel exploitation path that is not publicly disclosed at the time of our evaluation.

To avoid any potential bias, we do not update the execution boundaries after developing the \acp{PoC}, and we use all vulnerabilities to implement exploits following the same state-of-the-art attack scenarios as TensorAbuse.

\subsubsection{TensorAbuse Attacks}
Zhu et al.~\cite{zhu2025my} show that TensorFlow models can embed malicious behavior by abusing legitimate APIs, enabling system-level actions such as file-system access and network communication. This threat is triggered during the inference phase.

For this test, we use the \acp{PoC} provided in the accompanying repository of the original paper~\cite{zhu2025my}. The dataset includes four distinct exploits, each modeling a different attacker objective: \texttt{ExecuteCodeByInjection}, in which a malicious model creates a Python file that overrides imports of widely used libraries; \texttt{LeakFile}, which simulates data exfiltration by reading a sensitive file from the victim system (e.g., an SSH private key) and transmitting it to a remote attacker-controlled server; \texttt{LeakIP}, which triggers a request to a malicious server in order to disclose the victim's IP address; and \texttt{GetShell}, which appends a public key to the victim's authorized\_keys file, exfiltrates the username to a remote server, and simulates an attacker establishing SSH access using the injected key.

\ri detects all attacks during the inference phase; the results are reported in the first row of Table~\ref{tab:cve_tensorabuse_results}.

\begin{table}[t]
\centering
\small
\caption{Results for attacks based on TensorAbuse and framework vulnerabilities. Columns are attacker goals: EC (\texttt{ExecuteCodeByInjection}), LF (\texttt{LeakFile}), LIP (\texttt{LeakIP}), and GS (\texttt{GetShell}). \checkmark indicates detection. Due to vulnerability constraints, some attacks cannot be implemented (\texttt{NA}) or can only be partially implemented ($^{\dagger}$).}
\begin{tabular}{l ccccc}
\toprule
\textbf{Vulnerability} &  & \textbf{EC} & \textbf{LF} & \textbf{LIP} & \textbf{GS} \\
\midrule
TensorAbuse~\cite{zhu2025my} (TensorFlow)      & | & \checkmark & \checkmark & \checkmark & \checkmark \\
CVE-2024-3660~\cite{CVE-2024-3660} (Keras)    & | & \checkmark & \checkmark         & \checkmark        &  \checkmark        \\
CVE-2025-1550~\cite{CVE-2025-1550} (Keras)    & | & \checkmark & \checkmark & \checkmark & \checkmark \\
CVE-2025-8747~\cite{CVE-2025-8747} (Keras)    & | &  NA & NA         & NA         &  \checkmark$^{\dagger}$        \\
CVE-2025-9905~\cite{CVE-2025-9905} (Keras)    & | & \checkmark & \checkmark & \checkmark         & \checkmark \\
CVE-2025-9906~\cite{CVE-2025-9906} (Keras)    & | & \checkmark & \checkmark         & \checkmark        & \checkmark        \\
CVE-2025-12058~\cite{CVE-2025-12058} (Keras)   & | & NA & \checkmark$^{\dagger}$ & NA         & NA \\
CVE-2025-49655~\cite{CVE-2025-49655} (Keras)   & | & \checkmark         & \checkmark & \checkmark         & \checkmark \\
CVE-2025-32434~\cite{CVE-2025-32434} (Pytorch) & | & NA & NA & NA & \checkmark$^{\dagger}$ \\
\texttt{TFSMLayer} Abuse (Keras)   & | & \checkmark         & \checkmark & \checkmark         & \checkmark \\
\bottomrule
\end{tabular}
\label{tab:cve_tensorabuse_results}
\vspace{0.3em}
\end{table}

\subsubsection{Framework Vulnerabilities}
We focus on PyTorch and Keras which, together with TensorFlow covered by the previous experiment, span three of the most widely used ML frameworks~\cite{kaggle-survey-2022,jetbrains-python-survey-2024}. We identify vulnerabilities targeting these two frameworks by querying CVE.org using the keywords `keras' and `pytorch', and then selecting those that are compatible with our threat model.
We initially focus on vulnerabilities that enable arbitrary code execution and then extend the analysis to vulnerabilities that expose primitives compatible with TensorAbuse-style attacks.

Overall, we identify eight CVEs: seven affecting Keras and one affecting PyTorch. All Keras vulnerabilities manifest during the model loading phase, whereas the PyTorch vulnerability is triggered during inference (\texttt{forward} call). 
For each CVE, we take the official \acp{PoC} (which typically only spawn \texttt{/bin/sh} or write to mock files or SSH keys), whenever available, and extend them to implement complete attack scenarios inspired by TensorAbuse, reusing the same attacker-controlled servers.
When a vulnerability does not permit the full realization of a specific attack, we implement the closest achievable behavior. For example, when a vulnerability enables only arbitrary file writing, we use it to write to the authorized\_keys, simulating a subset of the \texttt{GetShell} attack.

In addition, we include PoCs implementing the same attack scenarios using \texttt{TFSMLayer} abuse, which was not publicly disclosed at the time of testing. This vulnerability affects Keras and manifests during model inference.

In total, the dataset for this test is composed of 27 \acp{PoC}. \ri successfully detects all attacks, confirming the effectiveness of its design, prioritizing actual exploit behavior over vulnerability-specific details.

\subsection{Large-Scale In-the-Wild Analysis (RQ2)}\label{sec:largehf}
We test our approach against a large number of models collected from the Hugging Face Hub. The goal is to assess the applicability of our approach to in-the-wild models, where false positives are more likely due to the lack of control over the samples w.r.t. controlled experimental settings.

We use the Hugging Face API to enumerate and download repositories containing PyTorch and Keras models. We select these two frameworks due to their widespread adoption in \ac{ML}~\cite{kaggle-survey-2022,jetbrains-python-survey-2024}, the high availability of publicly shared artifacts, and the large number of CVEs disclosed in 2025 (for Keras, see \S~\ref{sec:cveabuse}) or reliance on unsafe serialization formats (for PyTorch). We select repositories whose last update date falls within 2025 and further filter them using framework-related keywords (i.e., "keras" and "pytorch").

For each selected repository, we further filter the contained files by serialization format, considering \texttt{.bin}, \texttt{.pt}, and \texttt{.pth} files for PyTorch models, and \texttt{.h5} and \texttt{.keras} files for Keras models. While this filtering strategy may exclude some valid model artifacts, it is consistent with the goals of our evaluation, which aim to assess our approach on a large and diverse set of real-world model files rather than to analyze all models available on the platform. Moreover, due to computational constraints, we discard artifacts larger than 16~GiB and restrict the analysis to the loading phase.

In total, we analyze 40{,}162 repositories, from which we extract 88{,}789 artifacts. From the analysis we exclude models that rely on external dependencies and therefore cannot be fully executed (runtime dependencies are discussed in~\S~\ref{sec:limitations}). After this final filtering step, we retain 77{,}974 models ($\approx$88\%).
Among these, \ri flags 23 models as suspicious. We manually inspect all flagged cases and confirm that they exhibit malicious behavior, ranging from OS-level command execution to reverse shells, resulting in a false-positive rate of 0\%. We are in contact with Hugging Face, which is aware of this evaluation and its results.

\subsection{State-of-the-Art Comparison (RQ3)}\label{sec:sota}
To compare \ri against state-of-the-art approaches, we employ the same dataset recently used by \textsc{PickleBall}~\cite{pickleball}. This allows a direct comparison with \textsc{PickleBall} and its evaluation baselines: \textsc{ModelScan}~\cite{ModelScanProtectAI} (v0.8.7), \textsc{ModelTracer}~\cite{casey2024large}, and PyTorch's restricted unpickling mode (\texttt{weights\_only=True}, v2.8.0)~\cite{PyTorchSaveLoadRun}. For completeness, we also include Fickling~\cite{flickling} (v0.1.7), which provides analysis utilities and a decompiler for Pickle artifacts, as well as a runtime mechanism targeting \ac{ML} models that hooks the Pickle library to enforce security checks during loading.

We note that this evaluation focuses on Pickle-based model artifacts and that three of the compared approaches (\textsc{PickleBall}, \texttt{weights\_only}, and Fickling) operate exclusively on Pickle-based artifacts. In contrast, \ri is not specialized to any particular model format.

In the following, we discuss additional considerations regarding the experimental settings related to the dataset and threat model, followed by the corresponding results. We refer the reader to Appendix~\ref{app:cons} for further considerations on \texttt{weights\_only} and \textsc{ModelTracer}.

\mypar{Dataset Considerations}
The original dataset contains 253 benign models and 84 malicious models, for a total of 337 models~\footnote{The original paper reports 252 benign models; however, the official repository contains 253 samples, all of which we include in our analysis.}. Several malicious models are sourced from the repositories used to construct the MalHug dataset~\cite{zhao2024models}. MalHug also includes Keras models that are not supported by some of the defenses considered in the comparison (e.g., \textsc{PickleBall} and PyTorch’s \texttt{weights\_only} mode) and were therefore not included in the original \textsc{PickleBall}'s dataset.
We manually analyzed the dataset prior to experimentation. 
During this analysis, we identified three malicious models whose payloads rely on the \texttt{nt} module, an internal implementation module used by \texttt{os} on Windows platforms. As all experiments were conducted on Linux, we excluded these three models from our evaluation. 
We also identified a model labeled as malicious in the dataset (\texttt{twitter-roberta-base-sentiment.bin} from the \texttt{oceanhacktitude/tinymodel} Hugging Face repository\footnote{\url{https://huggingface.co/oceanhacktitude/tinymodel}}) that exhibited no malicious behavior; a detailed discussion is provided in Appendix~\ref{app:cons}. We therefore reclassified it as benign.
After these adjustments, the final dataset consists of 254 benign and 80 malicious models.

\mypar{Threat Model Considerations}
Among malicious models, 18 are of the form \texttt{eval("print('message')")} or similar variations. The \texttt{print} operation has no impact on the system an attacker aims to compromise under our threat model, and \texttt{eval} alone does not execute operating-system commands or access system resources. In other words, while \texttt{eval} can be abused for system compromise, its use alone does not constitute such a compromise, nor does it introduce capabilities beyond those already permitted by Pickle (i.e., Pickle already allows Python code execution).
One additional model contains strings commonly associated with exploit code, but consistently raises a \texttt{SyntaxError} across different environments before executing any payload and therefore does not represent an actual threat.
One model compiles malicious code and returns a code object that would require explicit execution. This is incompatible with our threat model, which assumes artifacts are loaded and used as \ac{ML} models.
Finally, another model only attempts to execute a Python script that is not included in the dataset, resulting in an exception. This behavior is not malicious per se, as it depends entirely on the contents of an external file that is not present and could be either benign or malicious. 

\begin{table}[t]
\centering
\caption{Comparison of detection outcomes under two threat models. We report, on the left, the results obtained using the ground truth defined by our threat model (see Section~\ref{sec:threat_model}), and, on the right, the results obtained using the original ground truth adopted by prior work (e.g., \textsc{PickleBall}).}
\label{tab:sota_comparison}
\resizebox{\linewidth}{!}{
\begin{tabular}{lcccc}
\toprule
\textbf{Tool} & \textbf{\#FP} & \textbf{\#FN} & \textbf{FPR} & \textbf{FNR} \\
\midrule
\textsc{ModelScan}~\cite{ModelScanProtectAI} & \pair{35}{16} & \pair{5}{7} &
\pair{12.7\%}{6.3\%} & \pair{8.5\%}{8.8\%} \\

\textsc{ModelTracer}~\cite{casey2024large} & \pair{1}{1} & \pair{17}{38} &
\pair{0.4\%}{0.4\%} & \pair{28.8\%}{47.5\%} \\

\texttt{weights\_only}~\cite{PyTorchSaveLoadRun} & \pair{117}{96} & \pair{\textbf{0}}{\textbf{0}} &
\pair{42.5\%}{37.8\%} & \pair{\textbf{0\%}}{\textbf{0\%}} \\

\textsc{PickleBall}~\cite{pickleball} & \pair{73}{52} & \pair{\textbf{0}}{\textbf{0}} &
\pair{26.5\%}{20.5\%} & \pair{\textbf{0\%}}{\textbf{0\%}} \\

Fickling~\cite{flickling} & \pair{117}{96} & \pair{\textbf{0}}{\textbf{0}} &
\pair{42.5\%}{37.8\%} & \pair{\textbf{0\%}}{\textbf{0\%}} \\

\textbf{\ri (ours)} & \pair{\textbf{0}}{\textbf{0}} & \pair{\textbf{0}}{21} &
\pair{\textbf{0\%}}{\textbf{0\%}} & \pair{\textbf{0\%}}{26.2\%} \\
\bottomrule
\end{tabular}}
\label{tab:tm_dual_confusion}
\end{table}

\begin{table*}[t]
\centering
\small
\caption{Generality of execution-boundary enforcement across evaluation dimensions.}
\label{tab:generality}
\setlength{\tabcolsep}{5pt}
\begin{tabular}{p{3cm} p{4cm} p{9cm}}
\toprule
\textbf{Dimension} & \textbf{Evaluated settings} & ~\textbf{Takeaway} \\
\midrule
ML frameworks &
PyTorch, Keras, TensorFlow &
Viewing models as variations within the capability space exposed by a framework is generally valid and not tied to a specific framework. \\
\addlinespace
Serialization formats &
Pickle, \texttt{.keras}, \texttt{.h5}, SavedModel &
Focusing on runtime behavior (effects on the host) rather than artifact structure generalizes across heterogeneous serialization formats. \\
\addlinespace
Framework versions &
Keras 3.8 (TF 2.17), Keras 3.12 (TF 2.20), PyTorch 2.5.1, Pytorch 2.9.1, TF 2.17 &
Expected actions are stable across framework versions: execution-boundary definitions required no changes, reducing maintenance burden. \\
\addlinespace
Framework--phase combinations &
PyTorch (loading, inference), Keras (loading, inference), TensorFlow (inference) &
The intuitions underlying our approach remain valid not only across frameworks but also across lifecycle phases.\\
\addlinespace
Attack classes &
Code injection, code reuse, framework vulnerabilities, etc. &
By focusing on \emph{what} a benign model \textit{should do} at runtime (allowlist-based approach) rather than on \emph{how} an attack is implemented, the approach remains effective across diverse exploitation techniques. \\
\addlinespace
In-the-wild models &
77{,}974 artifacts; 23 flagged, all confirmed malicious &
Execution boundaries derived from limited empirical data generalize well to real-world model artifacts beyond controlled settings, enabling low false-positive rates at scale. \\
\addlinespace
Dependency &
10 dependencies (beyond main framework libraries) installed during evaluation in~\S~\ref{sec:sota} (e.g., Ultralytics, Flair).&
After updating the orchestrator to correctly handle dependency initialization and imports, execution boundaries remained unchanged. This suggests that execution-boundary definitions are largely independent of specific dependency sets. \\
\bottomrule
\end{tabular}
\end{table*}

Accordingly, under our threat model, we reclassify these 21 models from malicious to benign, resulting in a dataset comprising 275 benign models and 59 malicious models. We acknowledge that prior work adopts a different labeling. For transparency, we report results under both threat models using the corresponding labels. While \ri is capable of monitoring actions such as writes to \texttt{stdout}, we deliberately do not modify execution boundaries when changing the ground truth, and instead consider the resulting additional false negatives to be an explicit design choice. 

\mypar{Results} Table~\ref{tab:sota_comparison} reports the results under the two threat models discussed. We present the number of false positives (\#FP), false negatives (\#FN), the false positive rate (FPR), and the false negative rate (FNR) for each evaluated tool.

Under the threat model this work assumes (reported on the left), \ri is the only approach that exhibits neither false positives nor false negatives. In contrast, approaches based on statically extracted signatures or allowlists (e.g., \textsc{ModelScan}, \textsc{PickleBall}, \texttt{weights\_only}, and Fickling) exhibit a substantial number of false positives while maintaining a low or zero false-negative rate. This behavior is expected: the malicious models in the dataset have been available for some time and exploit well-known vulnerabilities for which existing signatures and allowlists have already been updated. Finally, \textsc{ModelTracer} produces a single false positive (i.e., a model we reclassified from malicious to benign; see Appendix~\ref{app:cons} for further details); however, it suffers from false negatives due to its limited threat assumptions, as it restricts detection to network activity, process execution, and permission-modification events identified via four system calls and a few other \textit{commands}~\cite{pickleball}.

Under the threat model prior work assumes (reported on the right), \ri exhibits 21 false negatives, which are explicitly described in the \textit{Threat Model Considerations} paragraph. Regarding the other tools, the results trend remains largely unchanged: signature- and allowlist-based methods reduce their false positives while maintaining a low or zero number of false negatives. In contrast, \textsc{ModelTracer} exhibits an increase of exactly 21 false negatives.

\section{Discussion}
The experimental results covering three complementary research questions ((i) unseen exploits simulation over 31 PoCs, (ii) large-scale analysis of 77{,}974 in-the-wild artifacts, and (iii) state-of-the-art comparison over 334 models) confirm that our solution provides an effective way to detect malicious behavior in \ac{ML} model artifacts. The observed behavior provides evidence that, in the context of \ac{ML} model execution, it is possible to define execution boundaries such that (i) malicious behavior manifests as violations of lifecycle-specific execution boundaries, while (ii) benign models largely remain within them. 
Beyond raw detection accuracy, the results demonstrate that the proposed approach generalizes across multiple dimensions.
Table~\ref{tab:generality} summarizes the covered settings and the corresponding takeaways.

\mypar{Explainability and Diagnostic Value}
Although not the primary focus of this work, we observe that our approach can also support the understanding of malicious behavior. Boundary violations, by their nature, provide contextual information (e.g., violation type and accessed paths), offering actionable insights into how attacks manifest during execution. Within the scope of this work, this capability supported both the cross-checking of malicious model implementations against their observed boundary violations and the discovery of the \texttt{TFSMLayer} abuse described in \S~\ref{sec:cveabuse}.
\section{Limitations and Future Work}\label{sec:limitations}
\mypar{Anti-Dynamic Analysis}
Our method is potentially exposed to anti-dynamic analysis techniques, a well-studied class of evasions in the malware analysis literature~\cite{egele2008,chen2008towards,afianian2019malware}. 
The reference implementation, \ri, does not implement explicit countermeasures against such techniques and therefore does not claim resilience against an adaptive adversary seeking to evade runtime monitoring. 
At the same time, to the best of our knowledge, there is currently no evidence of \ac{ML} model artifacts in the wild employing anti-dynamic analysis techniques. We therefore view the absence of explicit anti-evasion mechanisms in \ri as a reflection of the current threat landscape rather than a design oversight.
Nevertheless, understanding whether and how classical anti-dynamic analysis techniques could be adapted to the structured execution environment of \ac{ML} models remains an important direction for future work.

\mypar{Execution Boundaries Maintenance}
In our experimental evaluation, we employed multiple versions of the same frameworks without requiring any redefinition or update of the execution boundaries, demonstrating their robustness. However, we cannot exclude the possibility that significant changes in framework implementations may eventually require updates to the associated boundaries. 
Given the allow-list nature of the expected action set, unreflected changes in framework behavior may manifest as false positives, with new benign actions flagged as malicious. Nevertheless, updating execution boundaries requires limited effort, mitigating the practical impact of this limitation. Moreover, this issue is not unique to dynamic analysis: static analysis techniques and rule-based scanners similarly require continuous maintenance to remain effective as frameworks evolve.

\mypar{Artifact Compatibility}
Model artifacts may be tied to specific versions of an \ac{ML} framework or of the Python interpreter, which can complicate their execution and their analysis. A possible mitigation is to maintain analysis environments supporting multiple frameworks and interpreter versions and to select the appropriate environment based on artifact metadata or on errors observed during preliminary execution attempts. In our evaluation of specific attacks and CVEs, we manually identified the appropriate framework versions and did not implement an automatic version-selection mechanism. We consider this an engineering effort rather than a conceptual limitation of our intuitions.

\mypar{Runtime Dependencies}
Dynamic analysis depends on the availability of runtime dependencies: if an artifact relies on external libraries, these must be available in the analysis environment for execution to proceed. Automated mechanisms could be explored to detect missing dependencies and provision them on demand within isolated environments. 
\section{Conclusion}\label{sec:conclusion}
\ac{ML} model sharing is increasingly part of the software supply chain, but existing defenses remain largely tied to specific formats, loading mechanisms, or known attack patterns. This work shows a complementary perspective: instead of reasoning about how malicious behavior is encoded inside an artifact, we reason about the effects that model execution produces on the host system.

We introduced \sys, a lifecycle-aware approach that secures \ac{ML} model execution by enforcing phase-specific execution boundaries over host interactions, and instantiated it in \ri, a syscall-based reference implementation. The key observation is that \ac{ML} models are not arbitrary software instances: they execute through a small number of well-defined lifecycle phases, and within each framework--phase pair their host interactions are structured enough to admit compact, reusable boundaries.

Our evaluation validates this across multiple dimensions. \ri detects all evaluated attacks across 31 \acp{PoC}, including framework vulnerabilities, code-reuse attacks, and inference-time threats, without relying on vulnerability-specific signatures. It scales to 77{,}974 artifacts from the Hugging Face Hub, flagging 23 models we confirmed as malicious, and outperforms state-of-the-art scanners under our threat model. These results indicate that lifecycle-aware dynamic analysis offers a practical and generalizable basis for securing \ac{ML} model execution.

More broadly, model scanning should not be treated solely as a static artifact-inspection problem. As \ac{ML} frameworks evolve and attacks increasingly abuse legitimate execution paths, defenses must monitor what model execution does to the host system. \sys is a step in this direction, showing that runtime effects, interpreted through the structure of the \ac{ML} lifecycle, can expose malicious behavior across formats, frameworks, and attack techniques.

%-------------------------------------------------------------------------------
% optional clearing of the page
\appendix

\bibliographystyle{plain}
\bibliography{bibliography}

\appendix

\section{Tracer Implementations}\label{app:tracing}
In this appendix, we survey concrete mechanisms that can be used to observe system-level interactions generated during \ac{ML} model execution and briefly discuss their respective trade-offs. These mechanisms represent possible implementation choices for the \textit{Tracer} (~\circnum{4}, see \S~\ref{sec:reference}).

\mypar{Ptrace-Based}
A natural instantiation consists of monitoring system calls issued by the process executing the \ac{ML} model. Since interactions with the host system (e.g., file access, network communication, process creation, or memory mapping) ultimately manifest as system calls, this approach provides fine-grained visibility into model and framework behavior. Traditional implementations rely on mechanisms such as \texttt{ptrace}~\cite{ptrace_man}, which allow a user-space monitor to intercept system calls and inspect their arguments. While expressive and precise, this approach may incur non-negligible overhead due to frequent context switches. This is the solution chosen for our reference implementation.

\mypar{eBPF-Based}
Extended Berkeley Packet Filter (eBPF)~\cite{sharaf2022extended} enables user-defined programs to execute safely within the kernel and to attach to observation points such as system calls, tracepoints, kernel functions, and \ac{LSM} hooks~\cite{wright2002linux}. Compared to ptrace-based tracing, eBPF can significantly reduce overhead by executing monitoring logic in-kernel and avoiding continuous user-kernel transitions.

\mypar{Seccomp-Based}
\texttt{seccomp}~\cite{seccomp_man} represents a restricted use of BPF programs, designed specifically to filter system calls invoked by a process. Seccomp filters are evaluated at the syscall boundary and can be installed statically or dynamically. However, unlike general-purpose eBPF tracing, seccomp policies offer limited semantic visibility. As a result, expressing lifecycle-aware execution boundaries or distinguishing between benign and unexpected uses of the same system call (e.g., file accesses to different paths) is challenging.

\mypar{Kernel Modules}
Kernel-level approaches include custom kernel modules and security frameworks such as \acp{LSM}~\cite{wright2002linux}. These mechanisms enable the enforcement of access control policies on operations such as file access, process creation, and network communication. However, they typically require system-wide configuration and administrative privileges, and their coarse granularity can complicate lifecycle-phase–specific enforcement.

\mypar{Existing Monitoring Tools}
Several host-based security and observability tools build on the mechanisms discussed above. For example, Falco~\cite{falco_project} and Sysdig~\cite{sysdig2025} leverage kernel modules or eBPF to provide runtime security analytics on Linux, while Sysmon exposes similar behavioral signals on Windows~\cite{microsoft_sysmon} and Linux~\cite{sysmon_linux}. Although effective for general-purpose threat detection, these systems are not designed for lifecycle-aware monitoring of \ac{ML} model execution.
\section{Additional Considerations on State-of-the-Art Comparison}\label{app:cons}
In the following, we present additional considerations regarding the model mislabeled as malicious in the \textsc{PickleBall}~\cite{pickleball} dataset, as well as \texttt{weights\_only} and \textsc{ModelTracer}, complementing the \textit{State-of-the-Art Comparison} in \S~\ref{sec:sota}.

\mypar{Considerations on a Mislabeled Sample} During our manual analysis of the \textsc{PickleBall} dataset, we identified one sample labeled as malicious (\texttt{twitter-roberta-base-sentiment.bin} from the \texttt{oceanhacktitude/tinymodel} Hugging Face repository\footnote{\url{https://huggingface.co/oceanhacktitude/tinymodel}}) that does not contain any malicious payload. Thereby, it cannot be considered malicious under either our threat model or that adopted by prior work. Consistently, the sample is not flagged as malicious by scanners supported by the Hugging Face platform.

Nonetheless, the sample is classified as unsafe by \textsc{PickleBall}, \texttt{weights\_only}, Fickling, and \textsc{ModelTracer}. The first three tools, which rely on statically constructed allowlists, do not include certain modules required by the model. For example, Fickling flags the model as unsafe due to its use of the \texttt{transformers} module, which is considered “outside the standard library.” In contrast, \textsc{ModelTracer} classifies the model as unsafe because it observes a \texttt{socket} system call during the loading phase. This system call is triggered by a \texttt{urllib} probe originating from a dependency of the \texttt{huggingface\_hub} module, which is imported indirectly during model loading as part of the model’s dependency chain. As a result, the observed system call is caused by the dependency chain rather than by the model itself. 

With respect to \ri, although it relies on system-call tracing, the orchestrator performs initialization steps that pre-load dependencies, including those that import the \texttt{urllib} module, thereby effectively isolating the lifecycle phase under analysis. This prevents the corresponding system call from being attributed to the model execution and being flagged as a violation.

\mypar{Considerations on \texttt{weights\_only}}
This mode is explicitly designed to ``\textit{limit the functions executed during unpickling to only those necessary for loading weights}''~\cite{PyTorchSaveLoadRun}. As a consequence, complete model artifacts that contain not only weight definitions but also model architecture, or that rely on third-party libraries, are not intended to be supported by this design. It is therefore expected that \texttt{weights\_only} blocks the loading of such artifacts. While, for the sake of comparison, we adopt the same evaluation metrics used in the original \textsc{PickleBall} work and classify these cases as false positives, this interpretation should be considered in light of the intended scope of the \texttt{weights\_only} mechanism.

\mypar{Considerations on \textsc{ModelTracer}} \textsc{ModelTracer} adopts a dynamic tracing approach. Specifically, it treats the occurrence of system calls and commands such as \texttt{socket}, \texttt{connect}, \texttt{execve}, \texttt{chmod}, \texttt{exec}, and \texttt{eval} during the model-loading phase as indicators of malicious behavior. As a dynamic analysis technique, \textsc{ModelTracer} faces challenges similar to ours in handling external dependencies. However, to avoid introducing additional noise or bias into its implementation, we evaluate \textsc{ModelTracer} exactly as released by the authors, without modifying its dependency-handling mechanisms.

\begin{figure}[!h]
\centering
\input{figures/fig_interaction_categories}
\caption{Categories of actions used in the reference implementation (\ri). Each category abstracts one or more system calls based on their semantics and arguments, and constitutes the basis for defining execution boundaries.}
\label{fig:categories}
\end{figure}

\begin{listing}[!h]
\centering
\begin{minipage}{\linewidth}
\begin{minted}[
    fontsize=\footnotesize,
    frame=none,
    bgcolor=gray!5,
    breaklines=true,
    breakanywhere=true
]{text}
ExecutionBoundaries:
  FILESYSTEM_READ:
    - path in PythonEnvironment
    - path in ModelFolder

  FILESYSTEM_MODIFY:
    - path in TorchTemporaryLoadingPaths

  DEVICE_ACCESS:
    - fd in {STDOUT, STDERR}

  PROCESS_CONTEXT:
    - allow all

  SYSTEM_GET_INFO:
    - allow time-related syscalls
\end{minted}
\caption{High-level execution boundaries for the loading phase of PyTorch models.}
\label{lst:high_level_boundary_pytorch_loading}
\end{minipage}
\end{listing}

\begin{listing}[!h]
\centering
\begin{minipage}{\linewidth}
\begin{minted}[
    fontsize=\footnotesize,
    frame=none,
    bgcolor=gray!5,
    breaklines=true,
    breakanywhere=true
]{text}
ExecutionBoundaries:
  FILESYSTEM_READ:
    - path in PythonEnvironment

  DEVICE_ACCESS:
    - fd in {STDOUT, STDERR}

  PROCESS_CONTEXT:
    - allow all
\end{minted}
\caption{High-level execution boundaries for the inference phase of Pytorch models.}
\label{lst:high_level_boundary_example}
\end{minipage}
\end{listing}

\begin{listing}[!h]
\centering
\begin{minipage}{\linewidth}
\begin{minted}[
    fontsize=\footnotesize,
    frame=none,
    bgcolor=gray!5,
    breaklines=true,
    breakanywhere=true
]{text}
ExecutionBoundaries:
  FILESYSTEM_READ:
    - path in PythonEnvironment
    - path in ModelFolder
    - path == /usr/lib/x86_64-linux-gnu/libc.so.6
    - path == /proc/meminfo
    - path == /etc/ld.so.cache
    - path == /etc/ld.so.preload
    - allow working-directory queries

  FILESYSTEM_EXEC:
    - path == /usr/lib/x86_64-linux-gnu/libc.so.6

  PROCESS_EXEC:
    - path == /usr/bin/uname
    - path == /usr/sbin/uname
    - path == /usr/local/bin/uname
    - path == /usr/local/sbin/uname

  DEVICE_ACCESS:
    - fd in {STDOUT, STDERR}

  PROCESS_CONTEXT:
    - allow all

  SYSTEM_GET_INFO:
    - allow system-information syscalls
    - allow time-related syscalls
\end{minted}
\caption{High-level execution boundaries for the loading phase of Keras models.}
\label{lst:high_level_boundary_keras_loading}
\end{minipage}
\end{listing}

\begin{listing}[!h]
\centering
\begin{minipage}{\linewidth}
\begin{minted}[
    fontsize=\footnotesize,
    frame=none,
    bgcolor=gray!5,
    breaklines=true,
    breakanywhere=true
]{text}
ExecutionBoundaries:
  FILESYSTEM_READ:
    - path in PythonEnvironment
    - path in ModelFolder
    - path == /usr/lib/x86_64-linux-gnu/libc.so.6
    - path == /etc/ld.so.cache
    - path == /etc/ld.so.preload

  FILESYSTEM_EXEC:
    - path == /usr/lib/x86_64-linux-gnu/libc.so.6

  PROCESS_EXEC:
    - path == /usr/bin/uname
    - path == /usr/sbin/uname
    - path == /usr/local/bin/uname
    - path == /usr/local/sbin/uname
    - allow memory-protection changes

  DEVICE_ACCESS:
    - fd in {STDOUT, STDERR}

  PROCESS_CONTEXT:
    - allow all

  SYSTEM_GET_INFO:
    - allow system-information syscalls
\end{minted}
\caption{High-level execution boundaries for the inference phase of Keras models.}
\label{lst:high_level_boundary_keras_inference}
\end{minipage}
\end{listing}

\begin{listing}[!h]
\centering
\begin{minipage}{\linewidth}
\begin{minted}[
    fontsize=\footnotesize,
    frame=none,
    bgcolor=gray!5,
    breaklines=true,
    breakanywhere=true
]{text}
ExecutionBoundaries:
  FILESYSTEM_READ:
    - path in PythonEnvironment
    - path in ModelFolder
    - path == /usr/lib/x86_64-linux-gnu/libc.so.6
    - path == /etc/ld.so.cache
    - path == /etc/ld.so.preload
    - path == /sys/devices/system/cpu/cpu0/tsc_freq_khz
    - path == /sys/devices/system/cpu/online
    - path == /proc/cpuinfo
    - path == /proc/sys/vm/overcommit_memory

  FILESYSTEM_EXEC:
    - path == /usr/lib/x86_64-linux-gnu/libc.so.6

  PROCESS_EXEC:
    - path == /usr/bin/uname
    - path == /usr/sbin/uname
    - path == /usr/local/bin/uname
    - path == /usr/local/sbin/uname
    - path == /usr/lib/x86_64-linux-gnu/libc.so.6
    - allow syscall mprotect

  DEVICE_ACCESS:
    - fd in {STDOUT, STDERR}

  PROCESS_CONTEXT:
    - allow all

  SYSTEM_GET_INFO:
    - allow system-information syscalls
\end{minted}
\caption{High-level execution boundaries for the inference phase of TensorFlow models.}
\label{lst:high_level_boundary_tensorflow_inference}
\end{minipage}
\end{listing}

%%%%%%%%%%%%%%%%%%%%%%%%%%%%%%%%%%%%%%%%%%%%%%%%%%%%%%%%%%%%%%%%%%%%%%%%%%%%%%%%
\end{document}